\documentclass[10pt]{iopart}

\usepackage{hyperref}
\hypersetup{
    colorlinks = true,
    allcolors = blue
}

\usepackage{amssymb,eurosym,mathrsfs,fancyhdr,CJK,multicol,graphics,indentfirst,color,bm,upgreek,booktabs,multirow,warpcol}

\usepackage{array, multirow}

\usepackage[outdir=./]{epstopdf}

\usepackage{graphicx}

\usepackage{csquotes}

\usepackage{caption}
\usepackage[symbol]{footmisc}

\begin{document}

\title[Reduction in benefits of total flux expansion on divertor detachment due to parallel flows]{Reduction in benefits of total flux expansion on divertor detachment due to parallel flows}

\author{M.Carpita$^1$\footnote[4]{ Author to whom any correspondence should be addressed.}, O. Février$^1$, H. Reimerdes$^1$, C. Theiler$^1$,\\ B. P. Duval$^1$, C. Colandrea$^1$, G. Durr-Legoupil-Nicoud$^1$, \\ D. Galassi$^1$, S. Gorno$^1$,  E. Huett$^1$, J. Loizu$^1$, L. Martinelli$^1$, \\ A. Perek$^1$, L. Simons$^1$, G. Sun$^1$, E. Tonello$^1$, C. Wüthrich$^1$ \\ and the TCV team\footnote[1]{ See author list of Reimerdes et al. 2022, Nucl. Fusion \href{https://iopscience.iop.org/article/10.1088/1741-4326/ac369b/meta}{62 042018}}}
\address{$^1$ EPFL-SPC, CH-1015 Lausanne, Switzerland}
\ead{massimo.carpita@epfl.ch}
\vspace{10pt}
\begin{indented}
\item[]June 2023
\end{indented}

\begin{abstract}

The Super-X divertor (SXD) is an alternative divertor configuration leveraging total flux expansion at the outer strike point (OSP). According to the \textit{extended 2-point model} (2PM), the key attractive features of the SXD are facilitated detachment access and control, but this is not always retrieved experimentally. However, parallel flows are not consistently included in the 2PM. In this work, the 2PM is refined to overcome this limitation: the role of total flux expansion on the pressure balance is made explicit, by including the effect of parallel flows. Consequentially, the effect of total flux expansion on detachment access and control is weakened, compared to predictions of the 2PM. This new model partially explains discrepancies between the 2PM and experiments performed on TCV, in ohmic L-mode scenarios, which are particularly apparent when scanning the OSP major radius Rt. In core density ramps in lower single-null (SN) configuration, the impact of Rt on the CIII emission front movement in the divertor outer leg - used as a proxy for the plasma temperature – is substantially weaker than 2PM predictions. Furthermore, in OSP radial sweeps in lower and upper SN configurations, in ohmic L-mode scenarios with a constant core density, the peak parallel particle flux density at the OSP is almost independent of Rt, while the 2PM predicts a linear dependence. Finally, analytical and numerical modelling of parallel flows in the divertor is presented. It is shown that an increase in total flux expansion can favour supersonic flows at the OSP. Parallel flows are also shown to be relevant by analysing SOLPS-ITER simulations of TCV.

\end{abstract}

\vspace{2pc}




\ioptwocol
\section{Introduction}

Power exhaust is a key challenge for the realisation of a magnetic confinement fusion reactor, such as tokamaks, as identified by the European roadmap for fusion energy \cite{Donne2018}. In a future power plant, large power losses from the confined plasma must be exhausted in a very narrow scrape-off-layer (SOL) region. The peak power density at the target, if unmitigated, are predicted to greatly exceed material limits \cite{Wischmeier2015}. Moreover, avoiding excessive long-term erosion on the reactor vessel components requires sufficiently low plasma target temperature \cite{Leonard2011}.

Diverted plasma configurations are employed for power exhaust, with the ability to support a large plasma temperature gradient between the confined plasma and the divertor targets. At sufficiently low electron temperature $T_e$, radiation by hydrogen and low-Z impurities becomes more efficient ($T_e\lesssim10~eV$), and the cross-sections for charge exchange ($T_i\lesssim5~eV$) and volumetric recombination ($T_e\lesssim1~eV$) increase, redistributing the exhausted power more isotropically, and transferring some of the plasma momentum and energy to neutrals \cite{Pitcher1997, Fenstermacher1999}. This allows the divertor to enter the detached regime,  greatly reducing the power and particle fluxes to the targets.

The lower-single-null (LSN) is currently the reference configuration for most operating tokamaks and is the chosen configuration for ITER \cite{Pitts2019}. Nonetheless, the extrapolation of this configuration to future reactors, with higher power and particle fluxes, cannot be safely assumed, in particular with respect to the integration of a detached divertor with a reactor-relevant core plasma. Alternative divertor configurations (ADCs) are, therefore, studied as potential solutions to this problem. ADCs' foreseen benefits include easier access to detachment, higher total radiated power capabilities, and better control over the location of the radiation front \cite{Theiler2017}. However, ADCs must be judged in the light of increased machine complexity, hence their assessment, through experiments and modeling, is crucial \cite{Reimerdes2020}. 

Among ADCs, one considered concept for future reactors is the Super-X divertor (SXD) \cite{Valanju2009}. Its main feature is an increase of the outer strike point (OSP) major radius $R_t$, which comes with an increased total flux expansion. The increase of $R_t$ increases the cross-sectional area of a flux tube $A_{\perp,t}$ (as the total magnetic field $B_{tot}$ is proportional to the inverse of the major radius $R^{-1}$) and, as a result, decreases the parallel power densities at the target, $q_{\parallel,t}$. For a constant grazing angle at the outer target, an increase in $R_t$ corresponds exactly to a linear increase of the target wetted area. The power density at the OSP then scales as $R_t^{-1}$, which has been demonstrated experimentally \cite{Theiler2017}. According to the \textit{extended 2-point model} (2PM) \cite{Stangeby2018, Lipschultz2016}, the key attractive features of the SXD are facilitated detachment access and control. However, in some cases, these predictions were neither retrieved experimentally \cite{Theiler2017, Petrie2013, Reimerdes2022} nor numerically \cite{Moulton2016}. In specific cases, it was argued that this disagreement with analytical predictions was explained by several possible effects, \textit{e.g.} target geometry \cite{Petrie2013}, neutral compression \cite{Fil2020, Meineri2023} and / or the divertor being in a sheath-limited regime \cite{Moulton2016}. However, a general understanding of the discrepancy has still not been obtained.

In this paper, the role of $R_t$ is discussed, both in terms of target conditions and for detachment access and control. Section \ref{sec:model} presents the 2PM, its predictions with respect to total flux expansion effects on detachment access and control and its modification to make the effect of parallel flows on the total pressure balance explicit, leading to predictions of weaker total flux expansion effects compared to the original ones. Section \ref{sec:TCVexp} presents SXD experiments on TCV tokamak \cite{Reimerdes2022} to investigate the role of $R_t$. Finally, in section \ref{sec:modelling}, the analytical and numerical modelling of parallel flows in the divertor is presented, showing that an increase in total flux expansion can favour supersonic flows at the OSP and that parallel flows are relevant by analysing SOLPS-ITER \cite{Wiesen2015} simulations of TCV. A summary and conclusions are presented in section \ref{sec:conclusions}.

\section{2PM extension accounting for parallel flows}
\label{sec:model}

The 2PM \cite{Stangeby2018, StangebyBook} is a reduced model which relates target quantities (electron temperature $T_{e,t}$, electron density $n_{e,t}$, parallel particle flux density $\Gamma_{t}$) with upstream control parameters for the SOL plasma, \textit{e.g.} the total plasma pressure $p_{tot,u}$ and the parallel power density $q_{\parallel,u}$ at the upstream location. These quantities pertain to one individual flux tube in the SOL and are linked together by momentum and power balances. The upstream location, labeled by $u$, is somewhat arbitrary, and can refer to the X-point location, the outer mid-plane (OMP), etc. It is usually taken as a stagnant location, \textit{i.e.} where $v_\parallel=0$. In the following, when needed, this location will be specified.

In the interest of completeness, in the 2PM, the parallel power density $q_\parallel$ is defined as the total parallel plasma power density, \textit{i.e.} in the simplest form (taking $n_e = n_i = n$ for densities and $T_e = T_i = T$ for temperatures)
\begin{equation}
    q_\parallel = (5nT + \frac{1}{2}m_inv_\parallel^2)v_\parallel + q_\parallel^{heat,cond}
    \label{Eq-totenflux}
\end{equation}
with $q_\parallel^{heat,cond}$ the total conducted heat flux density, $T$ the plasma temperature, $n$ the plasma density and $v_\parallel$ the parallel plasma velocity.

\subsection{2PM predictions for target quantities and their dependence on \texorpdfstring{$R_t$}{Rt}}
\label{subsec:2PM-og-pred}

The most general 2PM expressions for target quantities are reported by Stangeby in (15)-(17) of \cite{Stangeby2018}. These are equivalent to expressions obtained by Kotov and Reiter in \cite{Kotov2009} that were derived from the steady-state version of the equations solved by the 2D multi-species plasma fluid code B2.

These expressions are reported here, assuming the following simplifying hypotheses: (S-I) only hydrogenic ion species (\textit{i.e.} $n = n_e = n_i$) and no net current (\textit{i.e.} $v_\parallel=v_{e,\parallel}=v_{i,\parallel})$; (S-II) thermal equilibration is achieved along the flux tube (\textit{i.e.} $T=T_e=T_i$); (S-III) the plasma flow at the target is sonic (\textit{i.e.} $M_t = 1$, where $M=v_\parallel/c_s$ is the Mach number and $c_s=\sqrt{(T_e+T_i)/m_i}=\sqrt{2T/m_i}$ the  sound speed, and the subscript $t$ representing the target in what follows). Hypothesis (S-III) and its link to the total flux expansion effects are discussed in section \ref{subsec:MachEvo}. These assumptions, introduced for simplicity, can be easily relaxed and do not limit the following discussion. However, an additional assumption \textit{required} in the derivation of the following 2PM expressions is: (A-I) target quantities are evaluated at the sheath entrance (\textit{i.e.} $q_{\parallel,t} = q_{\parallel,se} = \gamma n_t T_t c_{s,t}$). Further details are provided in appendix \hyperref[app:2PMder]{A}. The expressions are
\begin{eqnarray}
    T_t^{2PM} = & \frac{8m_i}{\gamma^2} \cdot \frac{q_{\parallel,u}^2}{p_{tot,u}^2} \cdot \frac{(1-f_{cooling})^2}{(1-f_{mom-loss})^2} \label{Tt-2PM} \\ &\cdot \left ( \frac{R_u}{R_t} \right )^2 \nonumber \\
    n_t^{2PM} =& \frac{\gamma^2}{32 m_i} \cdot \frac{p_{tot,u}^3}{q_{\parallel,u}^2} \cdot \frac{(1-f_{mom-loss})^3}{(1-f_{cooling})^2} \label{nt-2PM} \\ &\cdot \left ( \frac{R_u}{R_t} \right )^{-2} \nonumber \\
    \Gamma_t^{2PM} =& \frac{\gamma}{8 m_i} \cdot \frac{p_{tot,u}^2}{q_{\parallel,u}} \cdot \frac{(1-f_{mom-loss})^2}{(1-f_{cooling})} \label{Gammat-2PM} \\ &\cdot \left ( \frac{R_u}{R_t} \right )^{-1} \nonumber
\end{eqnarray}
where $m_i$ is the ion mass, $\gamma\approx8.5$ the sheath heat transmission coefficient \cite{StangebyBook} and $R_{u/t}$ are the upstream and target major radii respectively. The power and momentum loss factors, $f_{cooling}$ and $f_{mom-loss}$, are
\begin{eqnarray}
    \frac{q_{\parallel,t}}{q_{\parallel,u}} \cdot \frac{R_t}{R_u} \equiv 1-f_{cooling} \label{fpwr-def}\\
    \frac{p_{tot,t}}{p_{tot,u}} \equiv 1-f_{mom-loss} \label{fmom-def}
\end{eqnarray}
and the total plasma pressure is
\begin{equation}
    p_{tot} = 2nT + m_inv_\parallel^2 = 2(1+M^2)nT
\end{equation}
The ratio $(R_u/R_t)$ in \eref{Tt-2PM}-\eref{Gammat-2PM} explicitly relates target quantities to total flux expansion. Both experiments and simulations \cite{Theiler2017, Petrie2013, Moulton2016} were done to test such specific explicit dependencies of target quantities on $R_t$, showing several discrepancies.

\subsection{Explicit dependence of \texorpdfstring{$f_{mom-loss}$}{f-momloss} on \texorpdfstring{$R_t$}{Rt} and the effective Mach number \texorpdfstring{$M_{eff}$}{Meff}}
\label{2pm-extended}

The loss factors $f_{cooling}$ and $f_{mom-loss}$ are lumped parameters accounting for a variety of complex physical processes \cite{Petrie2013, Moulton2016, Fil2020, Meineri2023}. These processes can be separated into two main groups: (1) volumetric sources and cross-field transport effects; (2) geometrical effects, related to flux tube cross-sections. This work focuses mainly on the latter, as they can be explicitly linked to total flux expansion effects, as shown in the following.

While $f_{cooling}$ relates only to processes pertaining to group (1), $f_{mom-loss}$ accounts also for geometrical effects. To show this, the total power and parallel momentum steady-state balances in a flux tube element are taken
\begin{eqnarray}
    \frac{1}{A_\perp}\partial_s (A_\perp q_\parallel) = S_{pwr} \label{id-pwrbal}\\
    \frac{1}{A_\perp} \partial_s(A_\perp mnv_\parallel^2) = - \partial_s (2nT) + S_{mom} \label{id-mombal}
\end{eqnarray}
where $s$ is a length coordinate along the flux tube and $S_{pwr/mom}$ are effective sources (or sinks) within the flux tube, respectively for power and momentum, related to processes pertaining to group (1). As in a flux tube $A_\perp \propto B_{tot}^{-1} \propto R$, rearranging \eref{id-pwrbal}-\eref{id-mombal} gives
\begin{eqnarray}
    \frac{1}{q_\parallel} \partial_s(q_\parallel) = \frac{S_{pwr}}{q_\parallel} - \frac{1}{R}\partial_s(R) \label{id-pwrbal2}\\
    \frac{1}{p_{tot}}\partial_s(p_{tot}) = \frac{S_{mom}}{p_{tot}} - \frac{\kappa}{R}\partial_s(R)
    \label{id-mombal2}
\end{eqnarray}
where $\kappa = mnv_\parallel^2/p_{tot} = M^2/(1+M^2)$ is the local ratio of dynamic and total pressure in the flux tube. Integrating \eref{id-pwrbal2}-\eref{id-mombal2} from upstream to target, rearranging and using \eref{fpwr-def}-\eref{fmom-def} gives
\begin{eqnarray}
    \frac{q_{\parallel,t}}{q_{\parallel,u}}\cdot \frac{R_t}{R_u} & =  exp\left(\int_u^t \frac{S_{pwr}}{q_\parallel}ds\right) \label{fpwr-def2 } \equiv \\
    ~ & \equiv 1-f_{cooling} \nonumber \\
    \frac{p_{tot,t}}{p_{tot,u}} & =  exp\left(\int_u^t \left[\frac{S_{mom}}{p_{tot}} - \frac{\kappa}{R}\partial_s(R)\right]ds\right) \equiv \label{fmom-def2} \\
    ~ & \equiv 1-f_{mom-loss}
\end{eqnarray}
It thus becomes apparent that $f_{mom-loss}$ includes geometrical effects, whereas $f_{cooling}$ does not. 
In literature, the influence of geometrical effects on $f_{mom-loss}$ was recognized,  but was not investigated in detail, as it was considered negligible or avoided for simplicity \cite{Stangeby2018, Moulton2016}.

To explicitly highlight the effect of total flux expansion on the total pressure variation, it is useful to rewrite \eref{fmom-def2} in a form similar to \eref{fpwr-def2 }. A constant $\kappa_{eff}$ is introduced, which satisfies 
\begin{equation}
    \int_u^t \frac{\kappa}{R} \partial_s(R) ds = \kappa_{eff} \int_u^t \frac{1}{R}\partial_s(R) ds
    \label{keff-def}
\end{equation}
$\kappa_{eff}$ is then the average of the ratio of dynamic to total pressure, weighted by the local relative variation of the flux tube area, between upstream and target. \eref{fmom-def2} then becomes
\begin{eqnarray}
    1-f_{mom-loss} &\equiv \frac{p_{tot,t}}{p_{tot,u}} = \label{fmom-R} \\ &= \left( \frac{R_u}{R_t} \right)^{\kappa_{eff}} exp\left(\int_u^t \frac{S_{mom}}{p_{tot}}ds\right) \nonumber
\end{eqnarray}
This equation now explicitly shows the effect of total flux expansion on the total pressure variation. It also shows the explicit dependence of $f_{mom-loss}$ on total flux expansion.

An additional quantity can be defined to substitute $\kappa_{eff}$ in \eref{fmom-R}, termed the effective Mach number $M_{eff}$
\begin{equation}
    M_{eff} = \sqrt{\frac{\kappa_{eff}}{1-\kappa_{eff}}} ~\leftrightarrow~\kappa_{eff} = \frac{M_{eff}^2}{1+M_{eff}^2} \label{Meff-def}
\end{equation}
From here, $M_{eff}$ will be used. Further insights on $\kappa_{eff}$ and $M_{eff}$, and their physical interpretation, are provided in appendix \hyperref[app:commentsphys]{B}.

\subsection{Consequence on target quantities scaling}
\label{subsec:cons-tar-sca}

\begin{figure}
    \centering
    \includegraphics[width = 1.0\columnwidth]{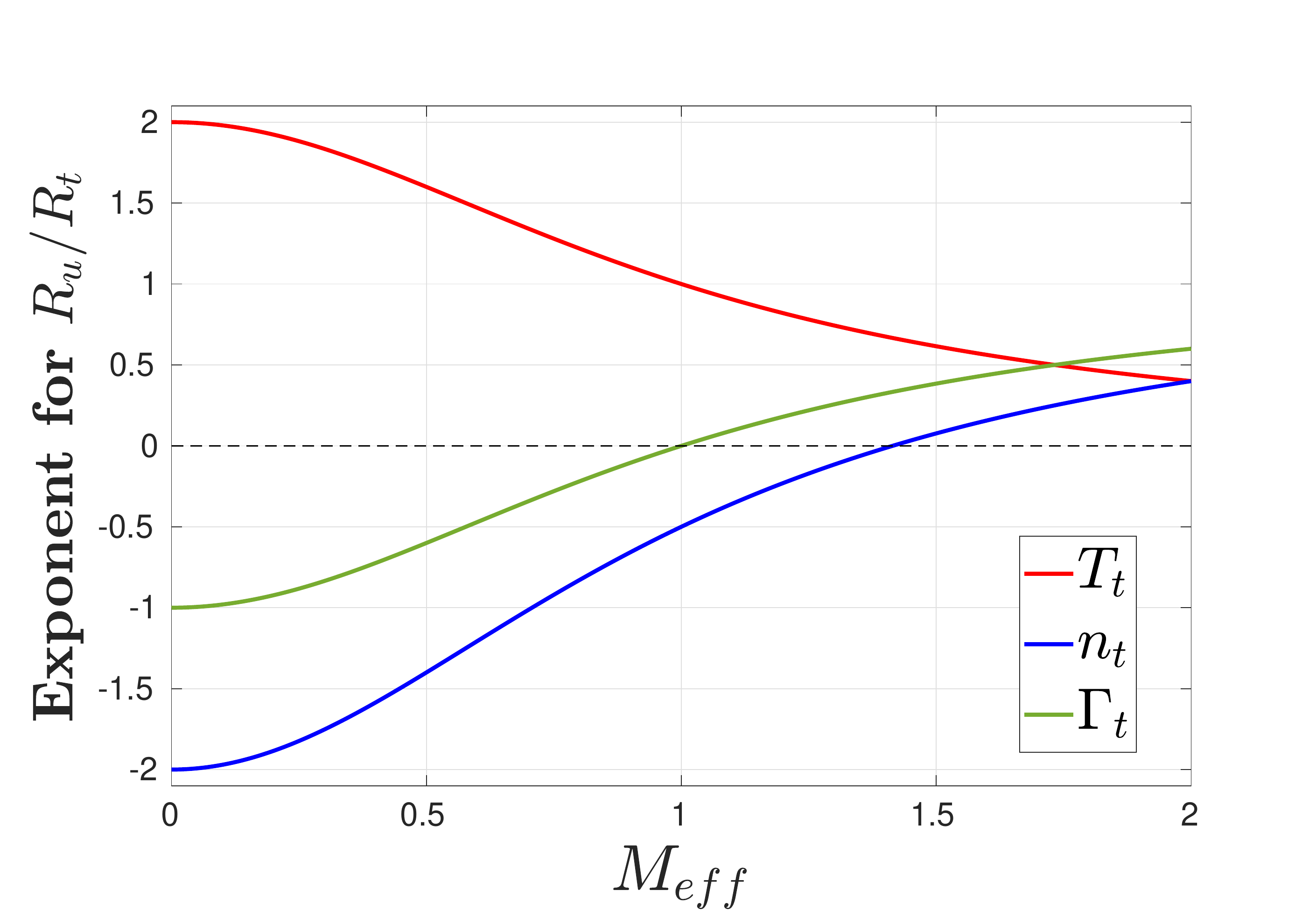}\\
    \caption{Exponents of $R_u/R_t$ for target temperature $T_t^{mod}$ (red), target density $n_t^{mod}$ (blue) and parallel particle flux density $\Gamma_t^{mod}$ (green), plotted against the effective Mach number $M_{eff}$.}
    \label{Fig:exponentsRt}
\end{figure}

The result obtained in \eref{fmom-R} is now considered together with \eref{Tt-2PM}-\eref{Gammat-2PM}. For the sake of clarity, the following notation is introduced
\begin{eqnarray}
    1-f_{cooling} \equiv (1 - f_{cooling}^S) \\
    1-f_{mom-loss} \equiv (1 - f_{mom-loss}^S) \\ ~~~~~~~~~~~~~~~~~~~~~ \cdot \left( \frac{R_u}{R_t} \right)^{\frac{M_{eff}^2}{1+M_{eff}^2}} \nonumber
\end{eqnarray}
so the newly defined factors $f_{cooling}^S$ and $f_{mom-loss}^S$ are accounting for the same physics, \textit{i.e.} volumetric sources and cross-field effects only. With this new definition of loss factors, \eref{Tt-2PM}-\eref{Gammat-2PM} then become
\begin{eqnarray}
    T_t^{mod} =& \frac{8m_i}{\gamma^2} \cdot \frac{q_{\parallel,u}^2}{p_{tot,u}^2} \cdot \frac{(1-f_{cooling}^S)^2}{(1-f_{mom-loss}^S)^2} \label{Tt-mod} \\ & \cdot \left ( \frac{R_u}{R_t} \right )^{2-\frac{2M_{eff}^2}{1+M_{eff}^2}} \nonumber \\
    n_t^{mod} =& \frac{\gamma^2}{32 m_i} \cdot \frac{p_{tot,u}^3}{q_{\parallel,u}^2} \cdot \frac{(1-f_{mom-loss}^S)^3}{(1-f_{cooling}^S)^2} \label{nt-mod} \\ & \cdot \left ( \frac{R_u}{R_t} \right )^{-2+\frac{3M_{eff}^2}{1+M_{eff}^2}} \nonumber \\
    \Gamma_t^{mod} =& \frac{\gamma}{8 m_i} \cdot \frac{p_{tot,u}^2}{q_{\parallel,u}} \cdot \frac{(1-f_{mom-loss}^S)^2}{(1-f_{cooling}^S)} \label{Gammat-mod} \\ & \cdot \left ( \frac{R_u}{R_t} \right )^{-1+\frac{2M_{eff}^2}{1+M_{eff}^2}} \nonumber
\end{eqnarray}
The dependence of target quantities on $R_u/R_t$ now varies with $M_{eff}$, figure \ref{Fig:exponentsRt}, and is weakened with increasing $M_{eff}$. The qualitative dependence of $\Gamma_t^{mod}$ and $n_t^{mod}$ on $1/R_t$ even reverses for $M_{eff}\geq 1$ and $\geq \sqrt{2}$, respectively. When $M_{eff}=0$, the dependence of target quantities on $R_u/R_t$ recovers the original ones in \eref{Tt-2PM}-\eref{Gammat-2PM}.

\subsection{Consequence on detachment window}
\label{subsec:cons-det-win}

It has been predicted that the Super-X configuration allows for an increased detachment front stability and control \cite{Reimerdes2020, Lipschultz2016, Cowley2022}, \textit{e.g.} a larger control parameters interval over which a detached regime can be achieved in the divertor while maintaining a tolerable impact on the core performance. This is a consequence of the negative parallel power density gradient that total flux expansion establishes towards the target, providing opposing response with respect to an upstream detachment front movement, \textit{i.e.} opposing radiation cooling. In terms of the operational window for detachment, Lipschultz \textit{et al.} (see (30) of \cite{Lipschultz2016}) provided an analytical estimate for the dependence of the detachment window on $B_{tot}\propto R_t^{-1}$

\begin{equation}
    \frac{\zeta_x}{\zeta_t} = \left[ \frac{R_t}{R_x} \right] ^\beta \label{detwin-0}
\end{equation}

where $\zeta_{x,t}$ are the values of a control parameter $\zeta=[p_u,f_I, P_{SOL}]$, that corresponds to the detachment front\footnote{The detachment front is intended here as the location where, ideally, all of the power loss occurs in a flux tube \cite{Lipschultz2016, Cowley2022}, separating a hot, attached upstream portion of the flux tube and a cold, detached downstream portion in front of the target.} being at the X-point or at the target, respectively. The three control parameters considered in this work are the upstream static pressure $p_u = 2n_u T_u$ (instead of $n_u$, as used in \cite{Lipschultz2016} - see appendix \hyperref[app:detwinder]{C}), the impurity fraction $f_I$ and the power entering the SOL in the flux tube of interest $P_{SOL}$. $R_{x,t}$ are the X-point and the target major radii, respectively. $\beta = [1,2,-1]$ is a specific exponent related to the considered control parameter.

The derivation of \eref{detwin-0} uses a momentum balance equivalent to the one in the 2PM and does not account explicity for any $p_u$  variation from upstream to target, \textit{i.e.} flux expansion effects and/or total pressure redistribution between dynamic and static contributions. When taken into account, the dependence of the detachment window on $B_{tot}\propto R_t^{-1}$ becomes

\begin{equation}
    \frac{\zeta_x}{\zeta_t} = \left[ \left(\frac{R_t}{R_x}\right)^{1-\frac{M_{eff}^2}{1+M_{eff}^2}} \cdot \frac{1+M_x^2}{1+M_t^2} \right] ^\beta \label{detwin-1}
\end{equation}

where the first factor in \eref{detwin-1} accounts for the total flux expansion and the second factor accounts for the total pressure redistribution. Further details on the derivation of \eref{detwin-1} are provided in appendix \hyperref[app:detwinder]{C}. The inclusion of total flux expansion and redistribution effects on total pressure reveals that the static pressure $p$ can include a gradient towards the target. In particular,  $p$ is proportional to the radiated power in the detachment front, as shown in \eref{RadQ2}. Consequentially, a negative gradient of the static pressure, as opposed to parallel power density, provides a positive feedback for the upstream movement of the detachment front and, hence, weakens the total flux expansion dependence of the detachment window.

\subsection{Summary of the effects of parallel flows on total flux expansion}

The impact of accounting for total flux expansion effects on momentum balance was shown and the following important points were highlighted:

\begin{itemize}
    \item The total pressure variation along a flux tube, see \eref{fmom-R}, can be linked explicitly to total flux expansion via $M_{eff}$, a lumped parameter characterising flows in the flux tube of interest.
    \item For negligible $M_{eff}$, this variation and its related effects are negligible. 
    \item Increasing $M_{eff}$ generally weakens the dependence on $R_t$ of target quantities, see \eref{Tt-mod}-\eref{Gammat-mod}, and detachment window, see \eref{detwin-1}, compared to predictions by the 2PM. In the case of \enquote{effective supersonic} flows ($M_{eff}\geq 1$), some dependencies also qualitatively reverse, starting with the particle flux.
    \item $M_{eff}$ depends on both the flow patterns in the flux tube and the geometrical design of the leg, in particular on the change of relative flux expansion along field lines, \textit{i.e.} $R^{-1}\partial_s(R)$, see \eref{keff-def} and \eref{Meff-def}. Two different divertor geometries, characterized by the same flow patterns and total flux expansion, can exhibit different behaviour with respect to their sensitivity to $R_t$. In appendix \hyperref[app:commentsphys]{B} this point is discussed in detail.
\end{itemize}

\section{SXD experiments in TCV and comparison with 2PM predictions}
\label{sec:TCVexp}

Experiments to investigate the SXD configuration are carried out in the \textit{Tokamak à Configuration Variable} (TCV) at EPFL \cite{Hofmann1994, Reimerdes2022}, testing the 2PM predictions presented in section \ref{subsec:2PM-og-pred}, regarding total flux expansion effects on detachment access and control. TCV is a medium-sized tokamak ($R_0 \sim 0.88~\mathrm{m}, ~B_0 < 1.45~\mathrm{T}, ~a \sim 0.25~\mathrm{m}$) with a highly elongated open vessel structure and a set of 16 independently-powered poloidal field coils, allowing for unique shaping capabilities that can generate many divertor configurations. The almost complete coverage of the vessel surfaces with graphite tiles allows for flexible placement of particle and power loads.

\subsection{Key diagnostics and experimental approach}

\begin{figure}[h!]
\centering
  {\includegraphics[clip,width=0.293\textwidth]{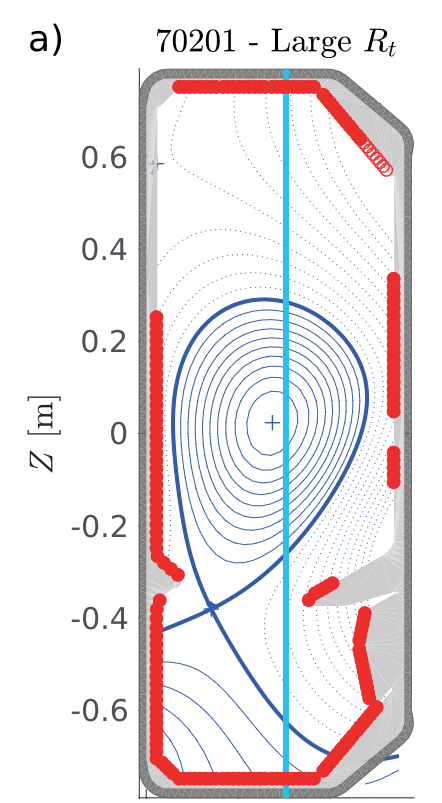}}
  {\includegraphics[clip,width=0.293\textwidth]{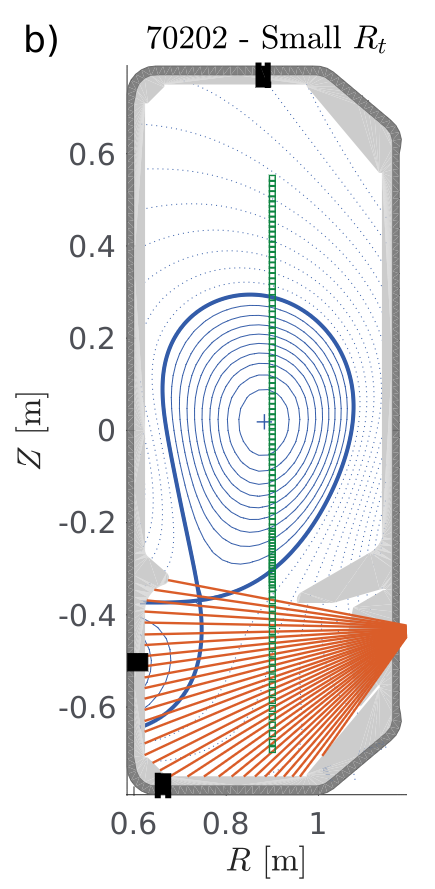}}
  \caption{Examples of baffled geometries used in the experimental work (large and small $R_t$). (a) The red dots indicate the position of wall-embedded Langmuir probes, while the cyan line indicates the FIR chord used for the feedback control of fuelling. (b) The black rectangles indicate the poloidal location of fuelling valves, the orange lines indicate the lines of sight of the DSS and the green dots indicate Thomson scattering measurement locations (intercepts between the laser and spectrometer lines of sight).}
  \label{fig:Exp_geometry}
\end{figure}

Different plasma geometries, characterized by varying OSP major radius $R_t$, are employed in this study, figure \ref{fig:Exp_geometry}. A set of polycrystalline graphite tiles, characterized by a longer structure on the low-field side compared to the high-field side (SILO baffles), is also employed in some experiments. They are designed to increase divertor closure, whilst maintaining good compatibility with alternative divertor configurations \cite{Reimerdes2021, Fevrier2021}.

$\mathrm{D}_2$ fuelling can employ alternative valves either on the floor, the inner wall or the ceiling of the vessel, figure \ref{fig:Exp_geometry}\textcolor{blue}{b}, allowing, among other things, to test the possible impact of fuelling locations on the results. The flow rates are feedback controlled and can be adjusted according to the line-averaged density $\langle n_e \rangle$ measurements by a far-infrared (FIR) interferometer, along a vertical chord, figure \ref{fig:Exp_geometry}\textcolor{blue}{a}. Density and temperature measurements in the core and across the separatrix are also measured by Thomson scattering \cite{Hawke2017}, figure \ref{fig:Exp_geometry}\textcolor{blue}{b}. Thomson scattering measurements also allow to compute $\langle n_e \rangle$ in the core. Wall-embedded Langmuir probes (LP) \cite{DeOliveira2019} cover a large part of the poloidal perimeter of the vessel, figure \ref{fig:Exp_geometry}\textcolor{blue}{a}. These were operated with a triangular voltage sweep (from $-120$ to $80~\mathrm{V}$ at $\sim330~\mathrm{Hz}$ and $\sim990~\mathrm{Hz}$ frequencies), in order to obtain temperature measurements as well as particle flux. Details on their analyses are provided in \cite{Fevrier2018}. The orange lines in the right panel show the lines of sight of a divertor spectroscopy system (DSS) \cite{Verhaegh2017}. Line radiation and their distributions are also obtained from a system of filtered optical cameras, MANTIS \cite{Perek2019}, that provide 2D poloidal map inversions of the emissivity for selected radiating spectral lines. This work focuses, in particular, on the CIII ($465.8~\mathrm{nm}$) line emission to obtain emissivity profiles. In previous TCV studies, the CIII radiation front along a divertor leg (determined as the location where the emissivity profile along the outer leg drops by $50\%$ with respect to the peak) was shown to provide a convenient estimation of the detachment status of the divertor. Due to a strong dependency on the local electron temperature, the CIII radiation front is a reliable proxy to identify the low temperature region along the outer leg \cite{Fevrier2021, Harrison2017}. A system of 64 gold foil bolometers, then substituted with a new system of 120 channels \cite{Sheikh2022}, is used to obtain radiation emissivity maps across a TCV poloidal section, by tomographically inverting their line integrated chord intensities. Finally, LIUQE \cite{Moret2015} is used to reconstruct the magnetic equilibrium across the discharges.

Two different scenarios are explored in this work, both with a plasma current $I_p \sim 250 ~\mathrm{kA}$ and the ion $\nabla B$ drift directed away from the X-point into the core, to avoid accessing H-mode \cite{Theiler2017}. The first employs ohmically-heated L-mode core density ramps $\langle n_e \rangle \simeq[4.0\rightarrow10.0]\cdot10^{19}~\mathrm{m}^{-3}$ (corresponding to $f_g\simeq[0.20\rightarrow0.55]$, $f_g$ being the Greenwald fraction). The density ramp is performed separately for two LSN configurations with small and large $R_t$, respectively. SILO baffles are employed to increase divertor closure, that is expected to improve the match between the 2PM predictions and experimental results, according to SOLPS-ITER simulations of TCV \cite{Fil2020}. Fuelling is performed from either the floor, inner wall (IW) or ceiling valves. The second scenario employs ohmically-heated L-mode OSP target radius $R_t$ scans at constant density $\langle n_e \rangle \simeq 5.5\cdot10^{19}~\mathrm{m}^{-3}$ ($f_g\simeq0.31$). This scenario is repeated in both Lower-Single-Null (LSN) or Upper-Single-Null (USN) configurations, with either SILO baffles or without, and floor-only fuelling.

\subsection{Density ramps at constant \texorpdfstring{$R_t$}{Rt}}
\label{subsec:dens_ramps}

\begin{table*}[h!]
\centering
\begin{tabular}{|c|c|c||c|c|c|}
\hline
\textit{Shot} & \mbox{$R_t$} & \textit{Fuel.} & $(1/B_{tot})^{OSP}~(T^{-1})$ & $L_{\parallel}^{OSP}~(m)$ & $f_{x,pol}^{OSP}$ \\ [2ex] \hline
70202         & Small         & IW                & 0.50                                    & 14.2                                    & 2.79                                 \\ \hline
70201         & Large         & IW                & 0.80                                    & 14.7                                    & 2.36                                 \\ \hline
63935         & Small         & Floor             & 0.50                                    & 13.6                                    & 2.82                                 \\ \hline
63917         & Large         & Floor             & 0.82                                    & 14.3                                    & 2.38                                 \\ \hline
63925         & Small         & Ceiling           & 0.50                                    & 13.8                                    & 2.83                                 \\ \hline
63934         & Large         & Ceiling           & 0.85                                    & 12.4                                    & 2.57                                 \\ \hline
\end{tabular}
\caption{Density ramps at constant $R_t$ - SOL geometry quantities at the OSP: inverse of the total magnetic field $(1/B_{tot}^{OSP}) \protect\propto R_t^{OSP}$, parallel connection length $L_{\parallel}^{OSP}$ (measured from the OMP, 5 mm from the separatrix) and poloidal flux expansion $f_{x,pol}^{OSP}$ (measured at 5 mm from the separatrix).}
\label{tab:SOL-related-quantities}
\end{table*}

Two values of $R_t$ are investigated during core density ramps: $R_t \simeq 0.62~\mathrm{m}$ (small $R_t$) and $R_t \simeq 1.03~\mathrm{m}$ (large $R_t$). When ramping the core density, the temperature in the divertor gradually reduces. Using the CIII front as proxy for low temperature region in the divertor, the 2PM prediction on temperature is tested in these experiments.

\begin{figure}
    \includegraphics[width = 0.95\columnwidth]{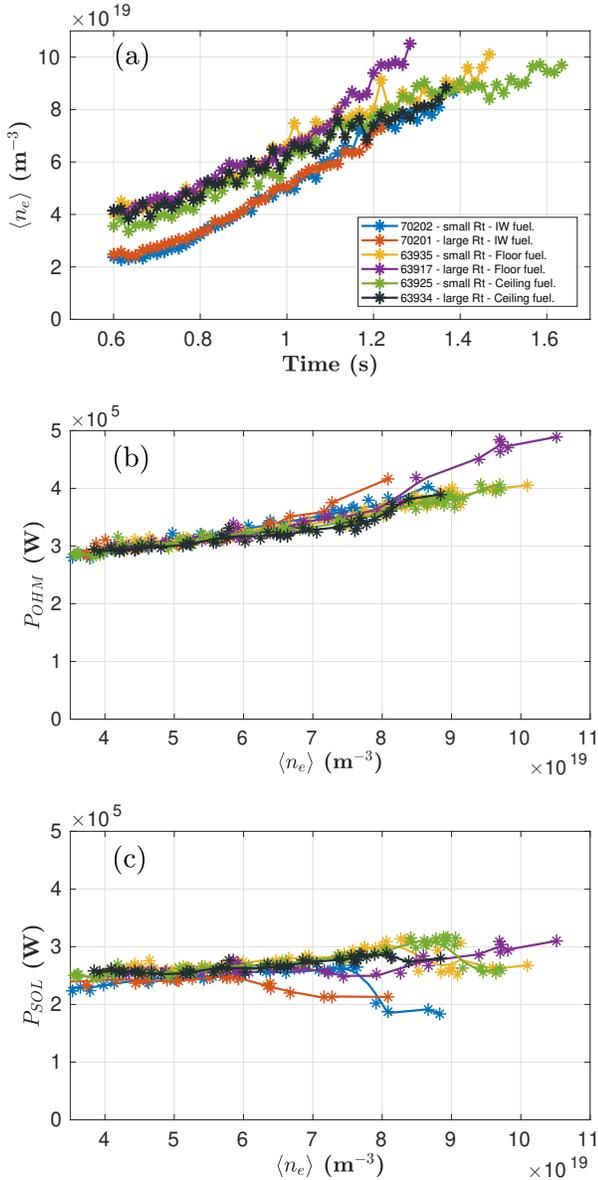}\\
    \vspace{-5mm}
    \caption{Density ramps at constant $R_t$ - (a) line-averaged density $\langle n_e \rangle$ variation in time; (b) ohmic power $P_{OHM}$ variation against $\langle n_e \rangle$; (c) power to the SOL $P_{SOL}$ variation against $\langle n_e \rangle$.}
    \label{Fig:dens_ramp_base}
\end{figure}

\begin{figure}[b]
    \hspace{-0.6cm}
    \includegraphics[width = 1.1\columnwidth]{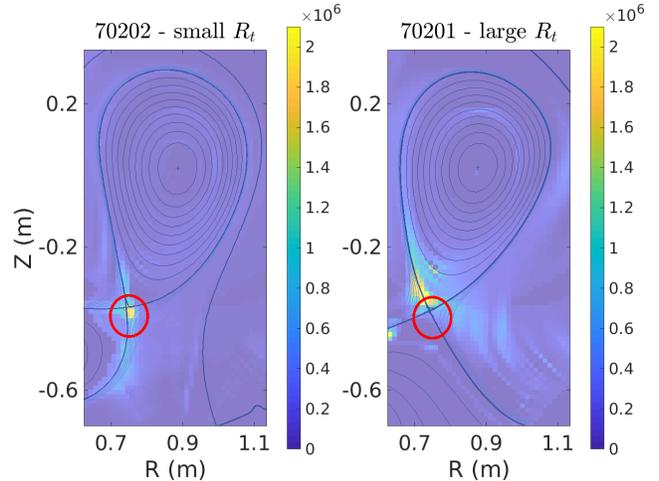}\\
    \vspace{-5mm}
    \caption{Density ramps at constant $R_t$, inner wall (IW) fuelling cases - Emissivity maps $(\mathrm{W}/\mathrm{m}^{3})$ at $\langle n_e \rangle = 6.75\cdot10^{19}~\mathrm{m}^{-3}$. The colormap is saturated at $2.1\cdot10^6~\mathrm{W}/\mathrm{m}^3$, to better highlight features of the emissivity maps away from the X-point. The red circle defines the $5 ~\mathrm{cm}$ radial area centered around the X-point, excluded from core radiation computation.}
    \label{Fig:emissIW}
\end{figure}

The discharges have a similar time evolution for $\langle n_e \rangle$ and input ohmic power $P_{OHM}$ dependence on $\langle n_e \rangle$, figures \ref{Fig:dens_ramp_base}\textcolor{blue}{a} and \ref{Fig:dens_ramp_base}\textcolor{blue}{b}. The power to the SOL, $P_{SOL}$, is defined as the difference between $P_{OHM}$ and the power radiated from the core, computed from bolometry, excluding a $5~\mathrm{cm}$ circular region centered around the X-point, figure \ref{Fig:emissIW}. $P_{SOL}$ dependence on $\langle n_e \rangle$ shows significant differences in cases with inner wall fuelling (up to $25\%$), figure \ref{Fig:dens_ramp_base}\textcolor{blue}{c}. Tomographic reconstruction of the emissivities for this fuelling location, figure \ref{Fig:emissIW}, suggests that this difference can be ascribed to increased radiation inside the confined plasma region at higher $\langle n_e \rangle$. Thomson scattering measurements (not shown) also show that the density and temperature in the core and near the separatrix remain comparable in all cases. Relevant SOL geometry quantities are reported in table \ref{tab:SOL-related-quantities}.

\begin{figure*}
    \centering
    \includegraphics[width=0.73\textwidth]{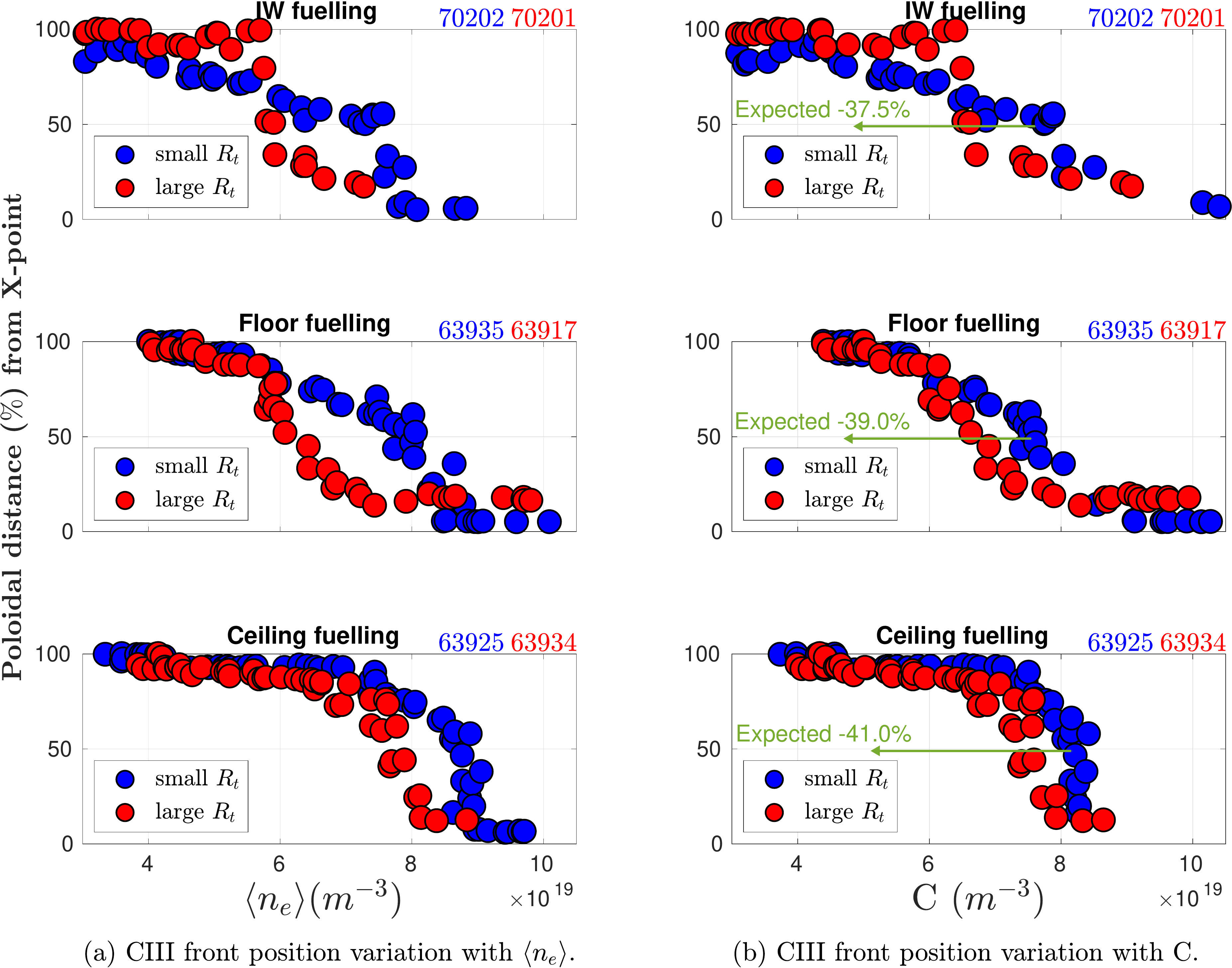}
    \caption{Density ramps at constant $R_t$ - CIII front position analyses from MANTIS along the outer leg - CIII front position is defined in terms of relative (\%) poloidal distance from the X-point, where 100\% is the target position. The expected shifts of large $R_t$ cases curves with respect to small $R_t$ cases are also plotted, computed according to \eref{eq:Tt-C}.}
    \label{fig:CIIIfront}
\end{figure*}

The CIII front movement at the outer leg against $\langle n_e \rangle$, taken from inversions of MANTIS measurements, is analysed to compare small and large $R_t$ configurations, figure \ref{fig:CIIIfront}\textcolor{blue}{a}. Similar results are obtained by the DSS (not shown). However, variations in $P_{SOL}$ and $L_{\parallel}^{OSP}$ can influence the CIII front behaviour between compared cases, as the front position is strictly related to the temperature in the divertor leg. According to the 2PM, the OSP target temperature $T_t$ (see \eref{Tt-2PM}, when changing the upstream control parameter from total pressure $p_{tot,u}$ to density $n_u$ \cite{Stangeby2018, StangebyBook}) is proportional to 
\begin{eqnarray}
    T_t^{2PM} & \propto \frac{1}{R_t^2} \cdot \frac{q^{10/7}_{\parallel,u}}{n_u^2 L_{\parallel}^{4/7}}
\end{eqnarray}
and taking
\begin{eqnarray}
    n_u \propto \langle n_e \rangle \label{approx_n&q} \\ q_{\parallel,u} \propto \frac{P_{SOL}}{\lambda_{sol,u}2\pi R_u B_{pol,u}/B_{tot,u}} \nonumber
\end{eqnarray}
one can write
\begin{eqnarray}
    T_t^{2PM} \propto \frac{1}{R_t^2} \cdot \frac{P_{SOL}^{10/7}}{\langle n_e \rangle^2 (L_{\parallel}^{OSP})^{4/7}} \label{Eq:C-def}
\end{eqnarray}

Note that this reasoning does not account for differences in other quantities between compared cases, such as: I) the geometrical location and features of the upstream location (\textit{e.g.} the scrape-off layer width $\lambda_{sol,u}$); II) in-out power sharing; III) the conducted-to-total power density ratio $f_{cond}$, IV) the ratio $n_u / \langle n_e \rangle$.

From \eref{Eq:C-def}, the parameter
\begin{equation}
    C\equiv\frac{\langle n_e \rangle (L_\parallel^{OSP}/L_\parallel^{ref})^{2/7}}{(P_{SOL}/P_{SOL}^{ref})^{5/7}}
\end{equation}
can be defined as a \textit{corrected} density. Plotting the CIII front movement against $C$ allows to consistently account for $P_{SOL}$ and $L_{\parallel}^{OSP}$ variations between compared cases, according to the 2PM. Here, $L_{\parallel}^{ref}=10~\mathrm{m}$ and $P_{SOL}^{ref}=2.5\cdot10^5~\mathrm{W}$ are considered. This is done in figure \ref{fig:CIIIfront}\textcolor{blue}{b}. From \eref{Eq:C-def}, the large $R_t$ configuration should see lower target temperatures for the same value of $C$. The CIII front movement from the target should thus happen at lower $C$ values for the higher $R_t$ cases. Given a specific front position obtained at values $C_{(small~R_t)}$ in the small $R_t$ cases, the expected reduced values for $C_{(large~R_t)}^{~expected}$ in the corresponding large $R_t$ cases can be computed as

\begin{equation}
    C_{(large~ R_t)}^{~expected} = C_{(small~ R_t)} \cdot \frac{R_t^{(small)}}{R_t^{(large)}}
    \label{eq:Tt-C}
\end{equation}

This is, however, not retrieved in the results shown in figure \ref{fig:CIIIfront}\textcolor{blue}{b}. For all the different fuelling cases, the variation in CIII front position with different $R_t$ is much weaker than predicted by the 2PM.

\begin{figure}[h]
    \includegraphics[width = 0.93\columnwidth]{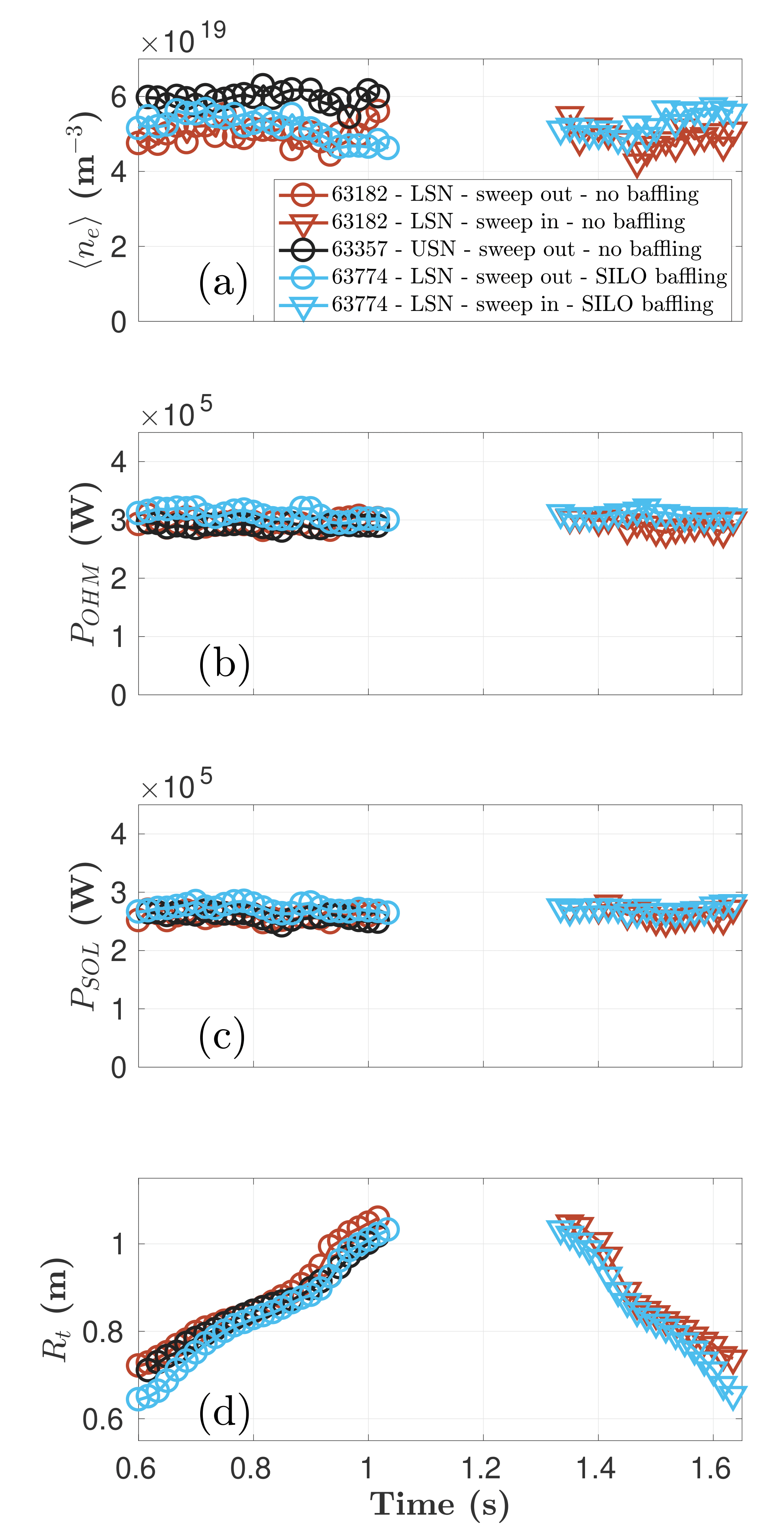}\\
    \vspace{-5mm}
    \caption{$R_t$ sweeps at approximately constant line-averaged density $\langle n_e \rangle$ - $\langle n_e \rangle$ (a), ohmic power $P_{OHM}$ (b), power to the SOL $P_{SOL}$ (c) and OSP major radius $R_t$ (d) variations in time. In between $\sim 1.05$ and $1.35~s$, the OSP is localised on a vessel segment for which a complete LPs coverage can not be achieved and, therefore, not of interest for the analyses and not reported here. For the USN case (black curve), only the outward sweep is available due to an early disruption.}
    \label{Fig:Rt_sweep_base}
\end{figure}

\begin{figure}
    \centering
      \includegraphics[clip,width=0.9\columnwidth]{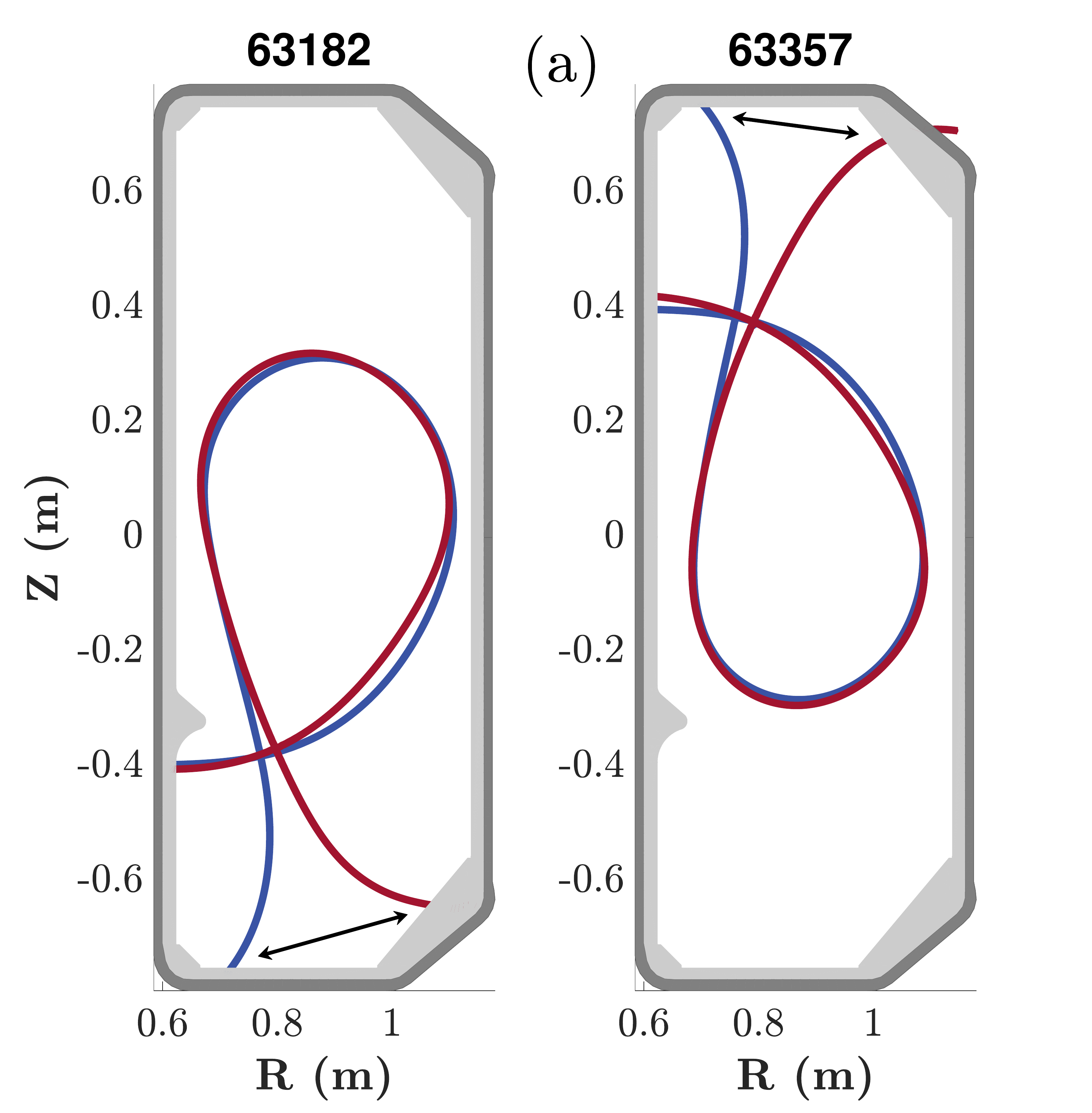}%
    
      \includegraphics[clip,width=0.95\columnwidth]{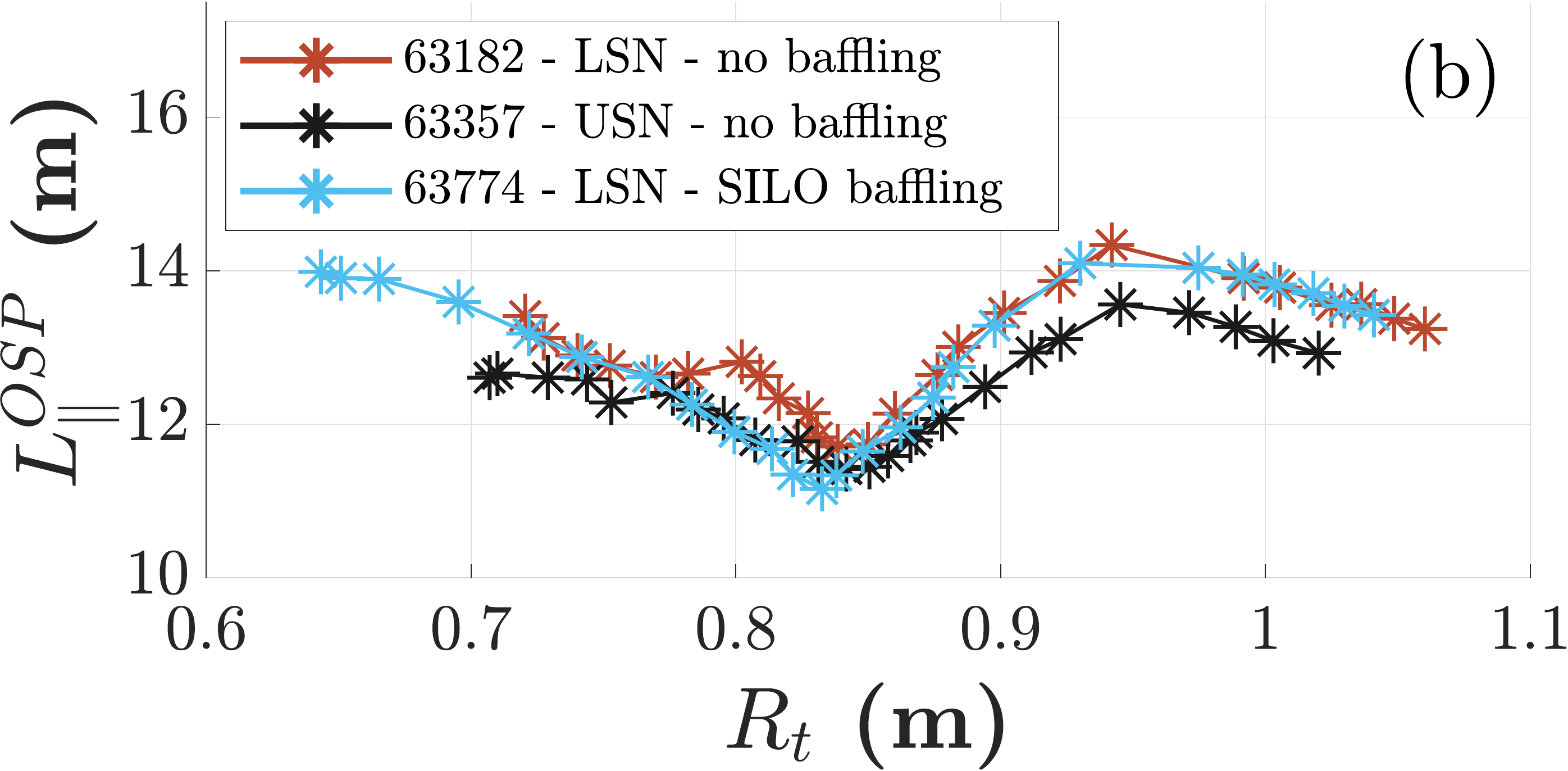}%
    \caption{$R_t$ sweeps at approximately constant line-averaged density $\langle n_e \rangle$ - (a) separatrix geometries for the unbaffled cases, showing the minimum and maximum $R_t$ achieved; (b) parallel connection length $L_{\parallel}^{OSP}$ (taken at the outboard midplane, $5 ~mm$ from the separatrix) variation against $R_t$.}
    \label{Fig:RtSweeps_divgeo}
\end{figure}

\subsection{\texorpdfstring{$R_t$}{Rt} scans at constant density}
\label{subsec:Rt_sweeps}

Here the opposite scenario is investigated by sweeping poloidally the OSP at constant core density, scanning a range of $R_t$ values ($R_t \simeq [0.7\leftrightarrow1.05]~\mathrm{m}$), with both outward and inward sweeps. When $R_t$ is modified, the target particle flux is also expected to vary according to the 2PM (see \eref{Gammat-2PM}). This prediction is tested in these experiments, using target LPs measurements.

During the strike-point sweeps, $\langle n_e \rangle$, $P_{OHM}$ and $P_{SOL}$ are kept approximately constant, figure \ref{Fig:Rt_sweep_base}, with observed variation of up to $10-20\%$. For all cases, $\langle n_e \rangle$ is always below $\sim 6.0\cdot10^{19} ~\mathrm{m}^{-3}$. At this density, for these experimental conditions, the CIII front in the outer leg remains close to the target and an attached state is maintained, as shown by the density ramps in \ref{subsec:dens_ramps}, figure \ref{fig:CIIIfront}\textcolor{blue}{a}. Thomson scattering measurements (not shown) show that the density and the temperature in the core and near the separatrix remain comparable across the strike-point sweeps. Figure \ref{Fig:RtSweeps_divgeo} plots the plasma geometry and $L_\parallel^{OSP}$ against $R_t$.

\begin{figure}[h!]
    \hspace{-0.5cm}
    \includegraphics[width = 1.15\columnwidth]{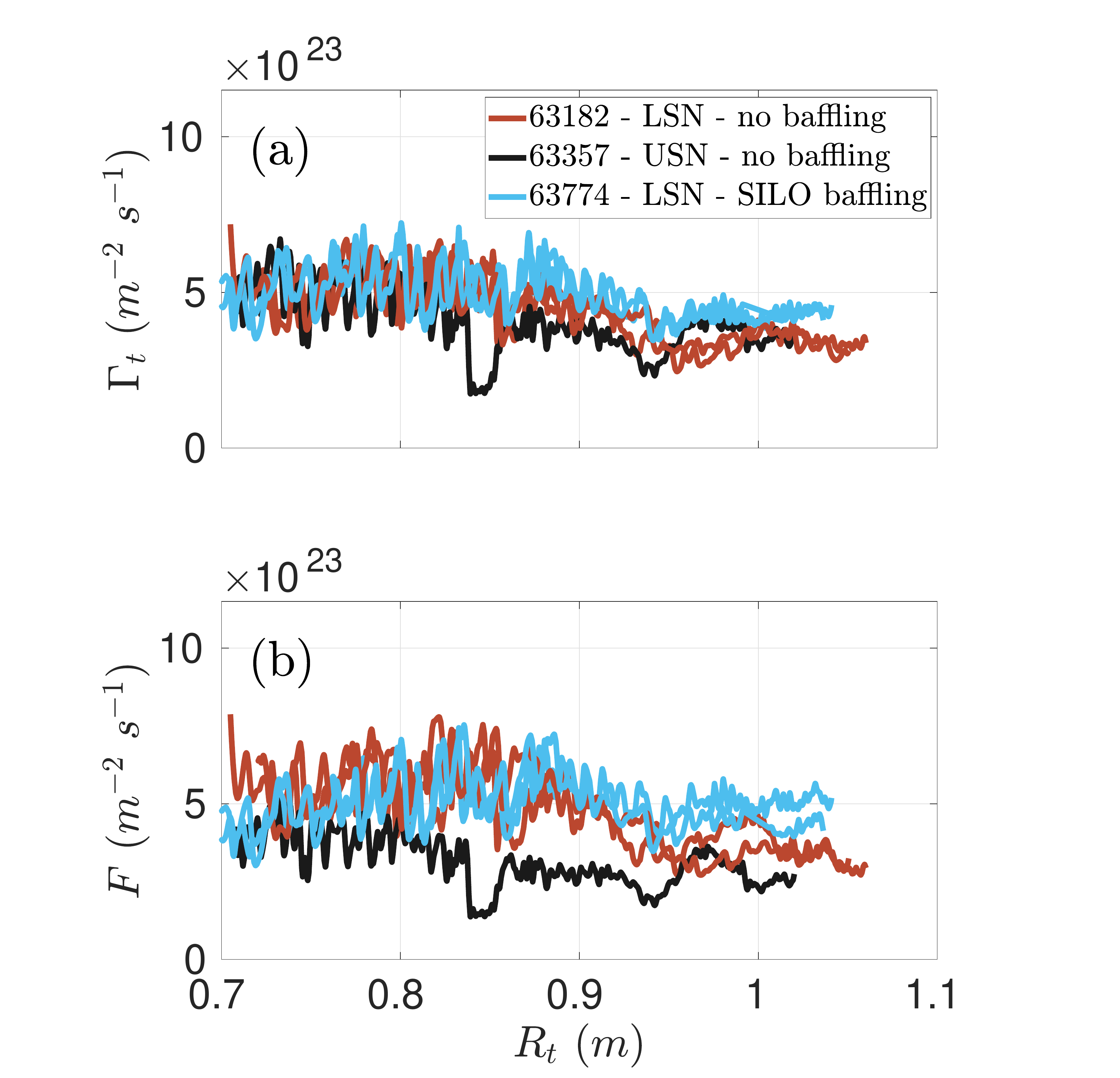}\\
    \vspace{-5mm}
    \caption{$R_t$ sweeps at approximately constant line-averaged density $\langle n_e \rangle$ - peak parallel particle flux density $\Gamma_t$ (a) and variable $F$ (b) against $R_t$. For shots in the LSN configuration (cyan and brown lines), two lines are reported representing the two sweeps performed (outward and inward).}
    \label{Fig:Rt_Scans_Gamma-vs-Rt}
\end{figure}

Figure \ref{Fig:Rt_Scans_Gamma-vs-Rt}\textcolor{blue}{a} shows the variation of the peak parallel particle flux density at the OSP $\Gamma_t$ against $R_t$. $\Gamma_t$ is taken from LPs measurements. However, variations in $\langle n_e \rangle$, $P_{SOL}$ and $L_\parallel^{OSP}$ can influence $\Gamma_t$ variations. According to the 2PM the OSP peak parallel particle flux density $\Gamma_t$ (see \eref{Gammat-2PM}, when changing the upstream control parameter from total pressure $p_{tot,u}$ to density $n_u$ \cite{Stangeby2018, StangebyBook}) is proportional to
\begin{equation}
    \Gamma_t^{2PM} \propto R_t \cdot \frac{n_u^2 L_\parallel^{4/7}}{q_{\parallel,u}^{3/7}} \propto R_t \cdot \frac{\langle n_e \rangle^2 (L_\parallel^{OSP})^{4/7}}{P_{SOL}^{3/7}} \label{eq:F-def0}
\end{equation}
Here, the same approximations (see \eref{approx_n&q}) employed in section \ref{subsec:dens_ramps} are used. From \eref{eq:F-def0}, the variable 
\begin{equation}
    F \equiv \frac{\Gamma_t ~(P_{SOL}/P_{SOL}^{ref})^{3/7}}{(\langle n_e \rangle/\langle n_e \rangle^{ref})^2 (L_\parallel^{OSP}/L_{\parallel}^{ref})^{4/7}} \propto R_t
    \label{eq:F-def}
\end{equation}
can be defined as a \textit{corrected} parallel particle flux density. Plotting $F$ against $R_t$ consistently accounts for $\langle n_e \rangle$, $P_{SOL}$ and $L_\parallel^{OSP}$ variations between compared cases, according to the 2PM. Here, $\langle n_e \rangle^{ref} = 5.5\cdot 10^{19} ~\mathrm{m}^{-3}$, $L_{\parallel}^{ref}=10~\mathrm{m}$ and $P_{SOL}^{ref}=2.5\cdot10^5~\mathrm{W}$ are considered. From \eref{eq:F-def}, $F$ is expected to increase linearly with $R_t$ which is, however, not observed in the experiments, figure \ref{Fig:Rt_Scans_Gamma-vs-Rt}\textcolor{blue}{b}. For all the different cases, the variation of $F$ with $R_t$ is much weaker than predicted by the 2PM. Significant discrepancies from the 2PM predictions, consistent with this result, are also observed for the integrated particle flux (not shown).

\section{Modelling of parallel flows in the divertor}
\label{sec:modelling}

SXD experiments in TCV, section \ref{subsec:dens_ramps} - \ref{subsec:Rt_sweeps}, showed much weaker total flux expansion effects than predicted by the 2PM. Parallel flows can potentially explain part of this discrepancy, section \ref{subsec:cons-tar-sca} - \ref{subsec:cons-det-win}. As a direct, reliable measurement of parallel flows was not available in the experiments, analytical and numerical modeling are presented in this section to assess if this effect can be significant in the experimental conditions.

\subsection{ Mach number evolution and possibility for supersonic flows}
\label{subsec:MachEvo}

The impact of parallel flows on total flux expansion effects increases with higher values of the  Mach number $M$ in the divertor, by definition of $M_{eff}$, \eref{keff-def} - \eref{Meff-def}. The evolution equation for $M$ along a SOL flux tube is presented here, obtained by combining particle and momentum balances. The simple case of a single hydrogenic ion species ($n_e=n_i=n$) is considered
\begin{eqnarray}
    (1-M^2) \partial_s (M) = & \frac{1+M^2}{nc_s}S_{par} 
    \label{eq-MachEvo} \\ & + \frac{M(1+M^2)}{c_s}\partial_s(c_s) \nonumber \\ & + A_\perp M \partial_s(\frac{1}{A_\perp}) -\frac{M}{m_inc_s^2}S_{mom} \nonumber
\end{eqnarray}
where $s$ is a length coordinate along the flux tube, increasing from upstream to target ($s=s_t$). $S_{par,mom}$ are effective sources/sinks in the flux tube, respectively for particles and momentum, related to volumetric sources and cross-field effects, see \eref{id-pwrbal}-\eref{id-mombal}. $c_s = \sqrt{T_e + T_i/m_i}$ is the local  sound speed. The derivation of \eref{eq-MachEvo} is shown in appendix \hyperref[app:MachEvo]{D}. It is important to note that \eref{eq-MachEvo} must satisfy the Bohm condition \cite{Bohm1949}, \textit{i.e.} $M\geq1$, at the target, as the target corresponds to the sheath entrance in this fluid model. Qualitatively, \eref{eq-MachEvo} shows that: 
\begin{itemize}
    \item four main drivers are responsible for $M$ variation along a field line: particle and momentum effective sources/sinks (both volumetric sources and cross-field effects),  sound speed $c_s$ variation and total flux expansion.
    \item the effect of these drivers is reversed when $M$ is lower or higher than $1$, \textit{i.e.} whether the plasma flow is subsonic or supersonic.
    \item a necessary (but not sufficient) condition for a supersonic transition is a change of sign of the right-hand-side in \eref{eq-MachEvo}.
\end{itemize}

Moreover, the constraint provided by the Bohm condition at the target allows to extract a sufficient (but not necessary) condition for the development of supersonic flows at the target. Taking a region $[s_t-\Delta s,s_t]$ before the target: \textit{if, in this region, the right-hand-side of} \eref{eq-MachEvo} \textit{is negative, then the flow must be supersonic}.

This case is interesting for the SXD configuration. Considering the ideal case where $S_{par,mom}$ and $\partial_s(c_s)$ are negligible across the region $[s_t-\Delta s,s_t]$, the right-hand-side of \eref{eq-MachEvo} is then negative in the same region in the outer leg, due to total flux expansion, and supersonic flows would arise. The idea that the convergent-divergent magnetic structure of a flux tube, such as in the outer leg of a SXD configuration, can favour supersonic flows at target has already been addressed before \cite{Bufferand2014,Riemann1995, Inutake2002}. The possibility for supersonic flows to arise at the OSP for the SXD configuration was already demonstrated numerically before \cite{Togo2019-1,Togo2019-2}. In consequence, $M_{eff}$ and parallel flow effects on total flux expansion are suggested to be potentially significant for the SXD configuration.

Moreover, when the other drivers are considered, for low target temperature (\textit{i.e.} $T_t \lesssim 5~eV$) as required in detached conditions, in front of the target:
\begin{itemize}
    \item $S_{par}$ is negative: at low temperatures the ionisation front moves away from the target and the the only effective particle sources/sinks will be radial transport\footnote{Here and in the following radial particle and momentum transport are considered negative contributions to $S_{par,mom}$. This is generally true for the hottest channels in the common flux region of the SOL.} and recombination that both make $S_{par}$ negative.
    \item $\partial_s(c_s)$ is negative.
    \item $S_{mom}$ is negative due to charge exchange, recombination and radial transport (thus $-S_{mom}$ will be positive).
\end{itemize}
In the outer leg of the SXD configuration, 3 out of 4 terms on the right-hand-side of \eref{eq-MachEvo} are therefore negative in the outer leg of a detached SXD configuration. This type of analysis can be also applied to other divertor configurations, even with negligible total flux expansion, and supersonic flows can arise for similar target conditions \cite{Marchuk2008, Takizuka2001, Ghendrih2011, Stangeby1991}. 

\subsection{SOLPS-ITER modelling of SXD experiments in TCV}
\label{subsec:SOLPSmodel}

\begin{figure}[h!]
    \includegraphics[width = 0.95\columnwidth]{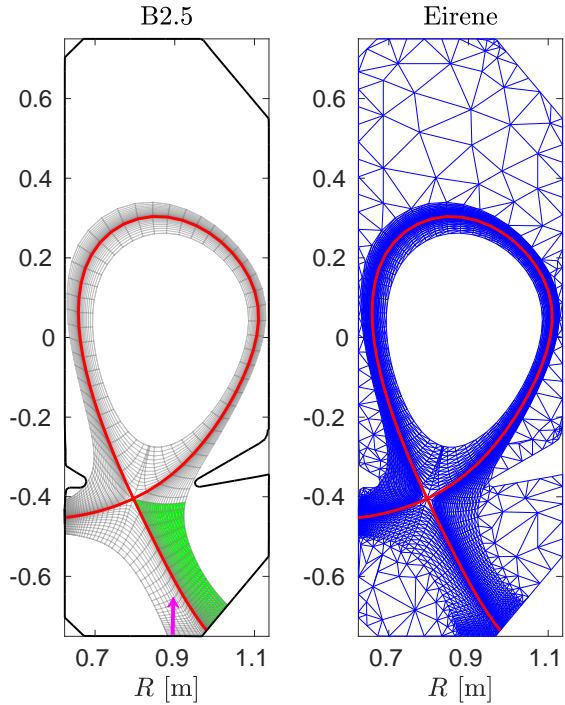}\\
    \vspace{-5mm}
    \caption{B2.5 and Eirene meshes for the SOLPS simulation. The pink arrow indicates the location for the fuelling. The green shaded area indicates the domain considered for the analyses of the outer leg.}
    \label{Fig:SOLPSgeom}
\end{figure}

A SOLPS-ITER simulation of TCV is used to study the patterns of parallel flows and $M_{eff}$ in the divertor region. SOLPS-ITER (B2.5-Eirene) is a transport code that couples the B2.5 multi-fluid solver with the kinetic Monte Carlo model neutral code Eirene \cite{Wiesen2015, Bonnin2016}. SOLPS-ITER is one of the most established SOL plasma simulators and it has been used for the design of the ITER divertor \cite{Kukushkin2011, Pitts2019}. The simulation discussed in this work was already presented in \cite{Wensing2019}, where details of the simulation setup are reported. The simulation features a baffled LSN geometry, figure \ref{Fig:SOLPSgeom}, with parameters typical of an ohmically-heated L-mode plasma in TCV, such as the experiments presented in section \ref{sec:TCVexp}. Drift effects are not included in this work, so radial transport is purely anomalous and incorporated by an artificial cross-field diffusion. The fuelling rate is varied to allow the analyses of different divertor conditions.

At the targets, a Dirichlet boundary condition satisfying the marginal Bohm criterion \cite{Bohm1949} is applied, \textit{i.e.} the parallel ion velocity at the sheath entrance is forced to match the plasma sound velocity (accounting for carbon impurities resulting from wall sputtering). This means that a Mach number $M=1$ at the target is imposed, excluding, \textit{a priori}, supersonic flows at the target (see section \ref{subsec:MachEvo}). This implies that the following evaluation of $M_{eff}$ is conservative: $M_{eff}$ could potentially have higher values in reality. 

To compute $M_{eff}$, the common flux region of the outer leg is considered in the simulation, taking as the upstream location the divertor entrance, figure \ref{Fig:SOLPSgeom}. This is also a conservative choice: the value of $M_{eff}$ usually has a minimum for a choice of upstream location which is close to the X-point (see appendix \hyperref[app:commentsphys]{B}). 

\begin{figure*}[t]
  \label{fig:Meff_A}{\includegraphics[clip,width=0.5\textwidth]{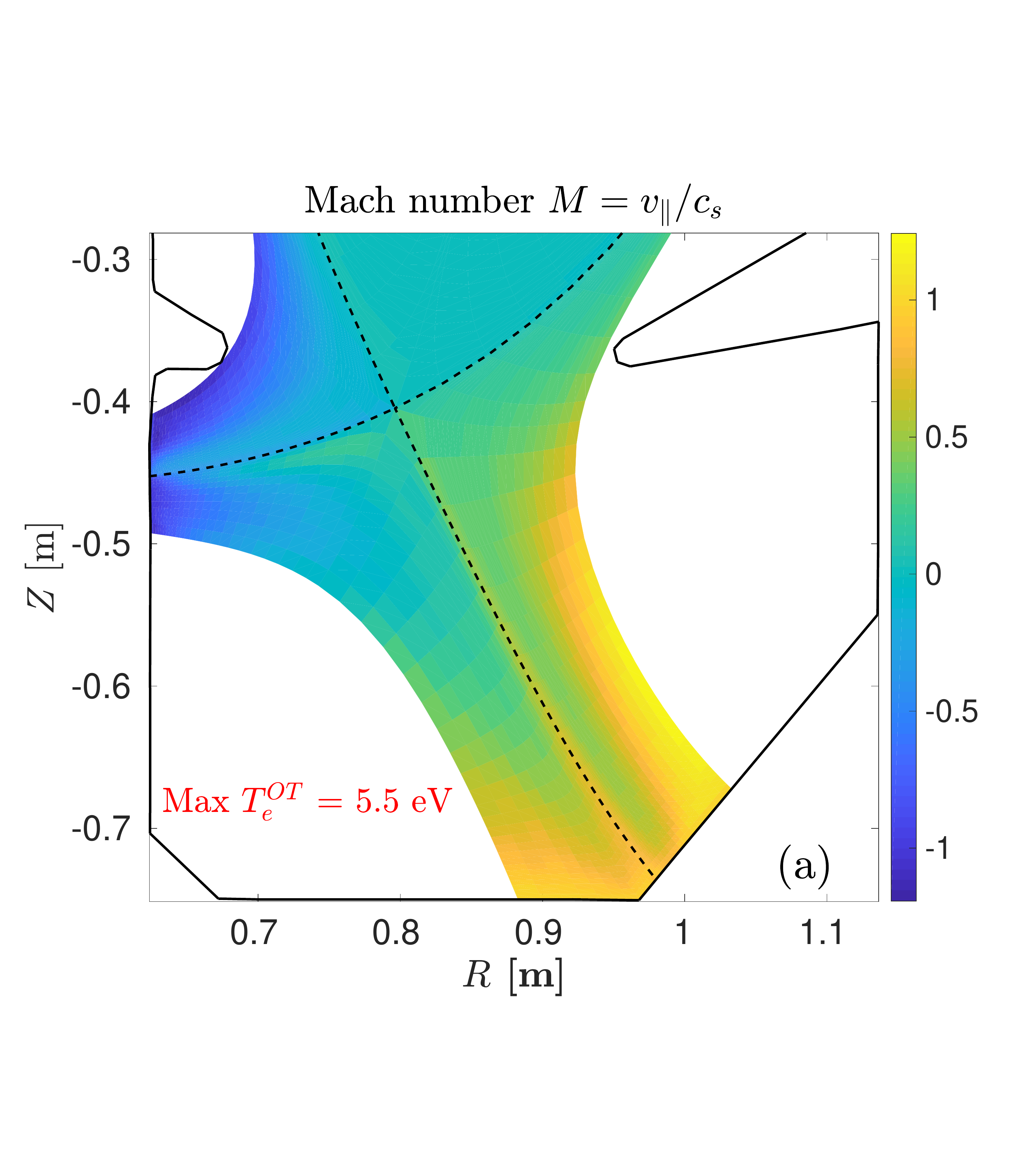}}
  {\includegraphics[clip,width=0.5\textwidth]{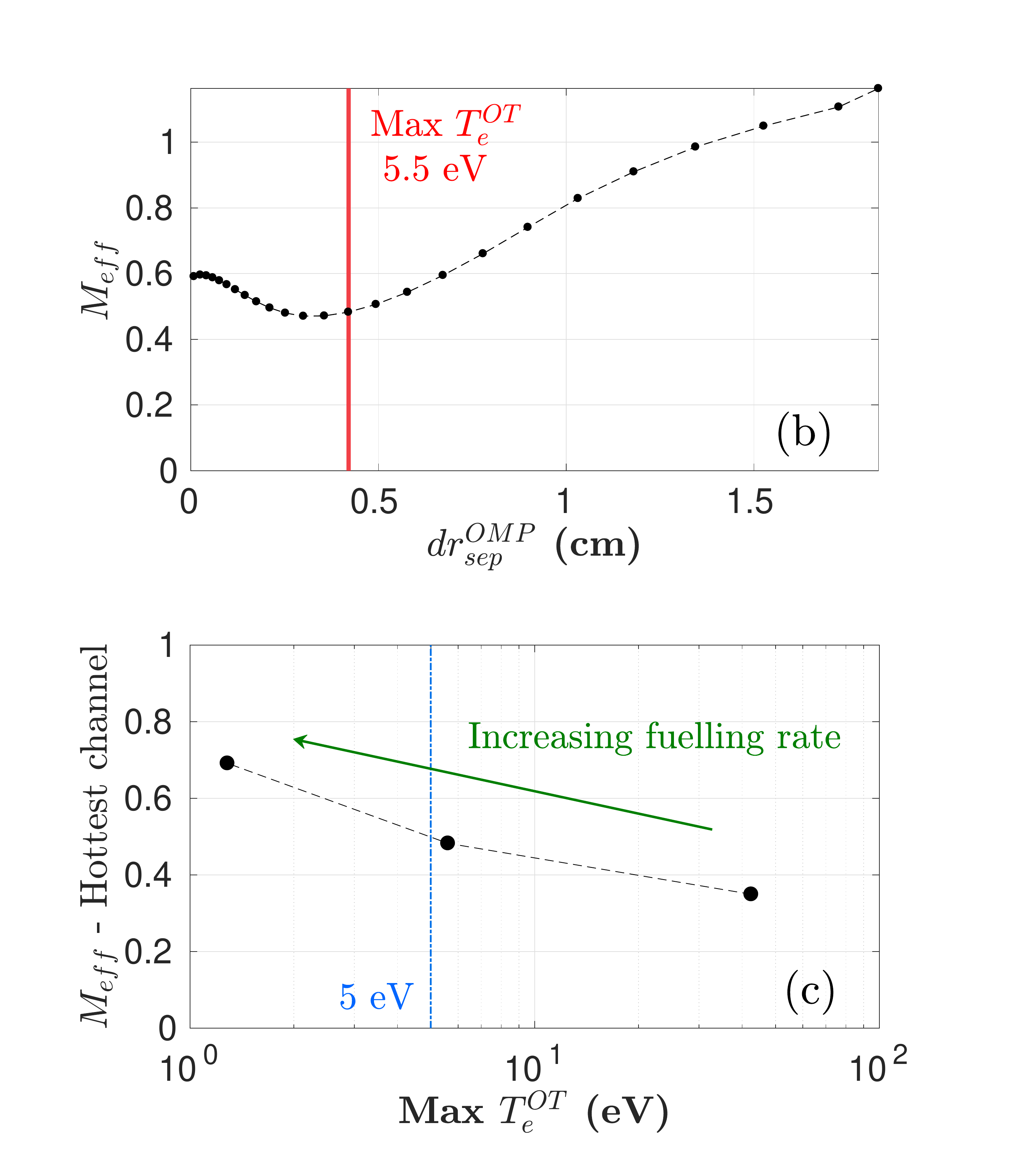}\label{fig:Meff_B}}
  \vspace*{-0.5cm}
  \caption{(a) Mach number $M=v_\parallel/c_s$ map in the divertor region, for the intermediate fuelling rate case, where $v_\parallel$ is the parallel velocity of the main plasma species $D^+$ and $c_s$ is the plasma sound speed, accounting for C impurities resulting from wall sputtering. (b) effective Mach number $M_{eff}$ for different flux tubes, in the case of intermediate fuelling rate, mapped against their radial distance from the separatrix at the OMP $dr_{sep}^{OMP}$. The flux tube with higher target temperature is indicated by the red vertical line. (c) effective Mach number $M_{eff}$ for the flux tube with higher target temperature ($\simeq~\mbox{Min }M_{eff}$), against $\mbox{Max }T_e^{OT}$.}
  \label{fig:Meff-sim}
\end{figure*}

For each flux tube in the analysed domain, $M_{eff}$ is evaluated according to \eref{keff-def} - \eref{Meff-def}. Its value varies both with the radial position of the flux tube, figure \ref{fig:Meff-sim}\textcolor{blue}{b}, and with divertor conditions, figure \ref{fig:Meff-sim}\textcolor{blue}{a}, as higher values are achieved for lower target temperatures. For the intermediate and higher fuelling rates, where divertor conditions are similar to the experiments presented in section \ref{sec:TCVexp}, $M_{eff}\geq0.5$ for all the flux tubes. In consequence, this SOLPS-ITER simulation suggests that $M_{eff}$ and parallel flow effects on total flux expansion are significant in these conditions, even with the conservative choices in the present analyses.

\section{Conclusions}
\label{sec:conclusions}

In this paper, the role of total flux expansion on the total pressure balance, neglected in the 2PM, is made explicit, by including the effect of parallel flows. This effect is quantified by a newly defined lumped parameter, the effective Mach number $M_{eff}$, characterising each flux tube. Its introduction allows to decouple geometrical effects from cross-field and sources/sinks effects in the momentum loss factor $f_{mom-loss}$. In consequence, 2PM target quantity expressions can be rewritten and their dependence on total flux expansion, through the ratio $R_u/R_t$, is now modified and varying with $M_{eff}$. For increasing $M_{eff}$, total flux expansion effects on target quantities is reduced and eventually qualitatively reversed, starting with the particle flux. The same modifications are applied to the detachment model by Lipschultz \textit{et al.}, showing how the dependence of the detachment window on total flux expansion weakens for increasing $M_{eff}$. Physically, this is ascribed to the fact that a negative static pressure gradient is established towards the target due to total flux expansion. 

Experiments on the SXD configuration are carried out in the TCV tokamak, testing 2PM predictions. These are all ohmically-heated L-mode discharges, in SN configuration, with $I_p \sim 250 kA$ and the ion $\nabla B$ drift directed away from the X-point. In core density ramps, featuring a baffled geometry and different fuelling locations, the CIII front movement in the outer leg, used as a proxy for the plasma temperature, shows variations with the outer strike point radius $R_t$ much weaker than 2PM predictions, especially when variations in $P_{SOL}$ and $L_\parallel$ are taken into account. In OSP sweeps, at approximately constant core density, the peak particle flux density at the OSP remains rather independent of $R_t$ variations, while a linear increase was predicted by the 2PM.

To understand if parallel flow effects can be significant in the experiments presented in this work, in the absence of experimental parallel flow measurements, both analytical and numerical modeling are employed. It is shown that supersonic flows, and therefore larger values of $M_{eff}$, are favoured in a SXD configuration due to the convergent-divergent magnetic structure of flux tubes in the outer leg. Moreover, the analyses of a SOLPS-ITER simulation of a baffled LSN geometry in TCV, with parameters typical of an ohmically-heated L-mode plasma, show that $M_{eff}\geq0.5$ in the outer leg, for divertor conditions similar to the ones present in the experiments, even with conservative choices in its evaluation. The modeling then suggests that parallel flows are, at least partially, causing the discrepancy between the 2PM predictions and the experiments. 

\section*{Acknowledgements}

This work has been carried out within the framework of the EUROfusion Consortium, via the Euratom Research and Training Programme (Grant Agreement No 101052200 — EUROfusion) and funded by the Swiss State Secretariat for Education, Research and Innovation (SERI). Views and opinions expressed are however those of the author(s) only and do not necessarily reflect those of the European Union, the European Commission, or SERI. Neither the European Union nor the European Commission nor SERI can be held responsible for them.


\appendix
\section[]{Derivation of 2PM expressions for target quantities}
\label{app:2PMder}

In this appendix the expressions \eref{Tt-2PM}-\eref{Gammat-2PM} are derived. To simplify the final expressions with respect to those reported in \cite{Stangeby2018}, it is assumed:
\begin{itemize}
    \item (S-I) only hydrogenic ion species (\textit{i.e.} $n = n_e = n_i$) and no net current (\textit{i.e.} $v_\parallel=v_{e,\parallel}=v_{i,\parallel})$.
    \item (S-II) thermal equilibration is achieved in the flux tube (\textit{i.e.} $T=T_e=T_i$).
\end{itemize}

An additional general assumption is needed:
\begin{itemize}
    \item (A-I) the target corresponds to the sheath entrance (\textit{i.e.} $q_{\parallel,t}=\gamma n_t T_t M_t \sqrt{2 T_t/m_i}$, where $M_t = v_{\parallel,t}/c_{s,t} = v_{\parallel,t}/\sqrt{2T_t/m_i}$ is the Mach number at the target and $\gamma$ is the sheath heat transmission coefficient). Note that, by Bohm condition at the sheath entrance, $M_t \geq 1$ must hold.
\end{itemize}

Introducing the standard definitions of power and momentum loss factors \eref{fpwr-def}-\eref{fmom-def} and using the above assumptions
\begin{eqnarray}
    (1-f_{cooling})q_{\parallel,u} R_u =&~ \gamma n_t T_t M_t \sqrt{\frac{2 T_t}{m_i}} R_t \label{step1-der} \\
    (1-f_{mom-loss}) p_{tot,u} = &~ 2(1+M_t^2) n_t T_t \label{step2-der}
\end{eqnarray}
where $p_{tot,t} = p_{tot,t}^e + p_{tot,t}^i = 2n_tT_t + m_i n_t v_{\parallel,t}^2 = 2 n_t T_t(1 + M_t^2)$.

The factor $n_t T_t$ is isolated in \eref{step2-der} and substituted into \eref{step1-der}, before isolating $T_t$ to obtain
\begin{eqnarray}
    T_t = & \frac{2 m_i (1 + M_t^2)^2}{\gamma^2 M_t^2} \cdot \frac{q_{\parallel,u}^2}{p_{tot,u}^2} \label{Tt-der} \\ &  \cdot \frac{(1-f_{cooling})^2}{(1-f_{mom-loss})^2} \cdot \left(\frac{R_u}{R_t}\right)^2 \nonumber
\end{eqnarray}
$n_t$ is then obtained from \eref{step2-der} and \eref{Tt-der}
\begin{eqnarray}
    n_t = &  \frac{\gamma^2 M_t^2}{4 m_i (1 + M_t^2)^3} \cdot \frac{p_{tot,u}^3}{q_{\parallel,u}^2} \label{nt-der} \\ & \cdot \frac{(1-f_{mom-loss})^3}{(1-f_{cooling})^2}\cdot \left(\frac{R_t}{R_u}\right)^2 \nonumber
\end{eqnarray}
Finally, $\Gamma_t$ is obtained as $n_t v_{\parallel,t} = M_t n_t\sqrt{2T_t/m_i}$
\begin{eqnarray}
    \Gamma_t = & \frac{\gamma M_t^2}{2 m_i (1 + M_t^2)^2} \cdot \frac{p_{tot,u}^2}{q_{\parallel,u}} \label{Gammat-der} \\ & \cdot \frac{(1-f_{mom-loss})^2}{(1-f_{cooling})}\cdot \left(\frac{R_t}{R_u}\right) \nonumber
\end{eqnarray}
Note that $\gamma = \gamma(M_t) \simeq 7.5 + M_t^2$ \cite{StangebyBook, Kotov2009}, therefore $\gamma_0 = \gamma(M_t = 1) \simeq 8.5$. The target quantities can then be rewritten, by grouping the terms directly depending on $M_t$ as factors of unitary value when $M_t = 1$
\begin{eqnarray}
    T_t = & \left( \frac{8.5 (1+M_t^2)}{2(7.5 + M_t^2) M_t} \right)^2 \cdot \frac{8 m_i}{\gamma_0^2} \cdot \frac{q_{\parallel,u}^2}{p_{tot,u}^2}  \\ & \cdot \frac{(1-f_{cooling})^2}{(1-f_{mom-loss})^2}  \cdot \left(\frac{R_u}{R_t}\right)^2 \nonumber \\
    n_t = & \left( \frac{8(7.5+ M_t^2)^2 M_t^2}{8.5^2 (1+M_t^2)^3} \right) \cdot \frac{\gamma_0^2}{32 m_i} \cdot \frac{p_{tot,u}^3}{q_{\parallel,u}^2} \\ & \cdot \frac{(1-f_{mom-loss})^3}{(1-f_{cooling})^2}\cdot \left(\frac{R_t}{R_u}\right)^2 \nonumber \\
    \Gamma_t = & \left( \frac{4(7.5 + M_t^2) M_t^2}{8.5 (1 + M_t^2)^2} \right) \cdot \frac{\gamma_0}{8 m_i} \cdot \frac{p_{tot,u}^2}{q_{\parallel,u}} \\ & \cdot \frac{(1-f_{mom-loss})^2}{(1-f_{cooling})}\cdot \left(\frac{R_t}{R_u}\right) \nonumber
\end{eqnarray}
These expressions recover \eref{Tt-2PM}-\eref{nt-2PM}-\eref{Gammat-2PM} when $M_t = 1$, that is hypothesis (S-III) in section \ref{subsec:2PM-og-pred}.

\section[]{Further comments and insights on total flux expansion effects on momentum balance and on the effective Mach number $M_{eff}$}
\label{app:commentsphys}

\noindent \textit{The synergy between parallel flows and total flux expansion on total pressure balance}

\bigskip

A short insight on the physical intuition behind the synergy between parallel flows and total flux expansion is provided here, highlighting the difference with the power balance.

Starting from \eref{id-mombal2} it is possible to notice that, contrary to the power balance expression \eref{id-pwrbal2}, the total flux expansion effect $-R^{-1}\partial_s(R)$ on the local total pressure variation is weighted by $\kappa = mnv_\parallel^2/p_{tot} = M^2/(1+M^2)$. In other words, the local flux expansion effect is re-scaled according to the local parallel flow conditions, in terms of $M$. In particular, for $M\ll1$, total flux expansion effects can be neglected.

The physical intuition is that the only component of the total pressure which is subject to the effect of locally varying cross-section in the flux tube is the dynamic pressure $mnv_\parallel^2$. This is because, in this work, the static pressure ($p=nT$) is considered isotropic\footnote{This assumption may be questionable in some conditions, and the anisotropy of pressure, especially for ions, in parallel and radial directions might play a direct role on total flux expansion effects \cite{Togo2019-2}.}, while the dynamic pressure is anisotropic, with a preferential direction along the flux tube. Mathematically, this is reflected in \eref{id-mombal} by the fact that dynamic pressure enters the balance via the divergence operator while the static pressure via the gradient operator.

\bigskip

\noindent \textit{Mathematical definition of $M_{eff}$, counter-intuitive values and its dependence on the upstream location}

\bigskip

From \eref{keff-def} - \eref{Meff-def}, $M_{eff}$ (or $\kappa_{eff}$) can be defined as: \textit{the value of $M$ (or $\kappa$) which, when constant from upstream to target, would provide the same total pressure variation $p_{tot,t}/p_{tot,u}$ due to total flux expansion}. Despite $\kappa = M^2/(1+M^2) \in [0,1)$, from \eref{keff-def} it is clear that $\kappa_{eff}$ can in principle take on any real value, due to the averaging process against $R^{-1}\partial_s(R)$. This reflects in $M_{eff} \rightarrow +\infty$ for $\kappa_{eff} \rightarrow 1^{-}$ or $M_{eff}$ assuming imaginary values for $\kappa_{eff} \notin [0,1]$ (see \eref{Meff-def}). Despite being counter-intuitive, this does not pose a direct problem to the mathematical formulation: $M_{eff}$ always enters the expressions presented in this work as $M_{eff}^2/(1+M_{eff}^2)=\kappa_{eff} \in \mathbb{R}$. The remaining problem is when $\kappa_{eff} \rightarrow \pm \infty$, that can happen for $R_u \rightarrow R_t$. In this case, in the expressions presented in this work, the indeterminate form $(R_u/R_t)^{\kappa_{eff}}$ would appear. This is a consequence of forcing the geometrical term in the total pressure variation $(1-f_{mom-loss})$ to take the form of a power of $(R_u/R_t)$ (see \eref{fmom-def2} - \eref{fmom-R}). However, this was necessary to maintain a simple form compatible with the algebraic expressions of the 2PM. 

Here, a pathological example is provided to discuss the meaning of infinite or imaginary values for $M_{eff}$, which could be difficult to understand in terms of the $M_{eff}$ definition provided above. This also shows how the $M_{eff}$ value depends on the upstream location. Consider a LSN geometry and focus on computing $M_{eff}$ for a flux tube in the outer divertor leg, varying the upstream location from the OSP to the OMP. A parallel length coordinate $s$ is defined, increasing from $s=s_{OMP}$ at the OMP to $s=s_t$ at the OSP. Assume that:
\begin{itemize}
    \item the Mach number is unitary between the X-point and the OSP and null elsewhere, that is $M=1\cdot \chi[s_x, s_t]$
    \item $R_x<R_t<R_{OMP}$, where $R$ is the major radius.
\end{itemize}

\begin{figure}[b]
    \centering
      \includegraphics[clip,width=0.75\columnwidth]{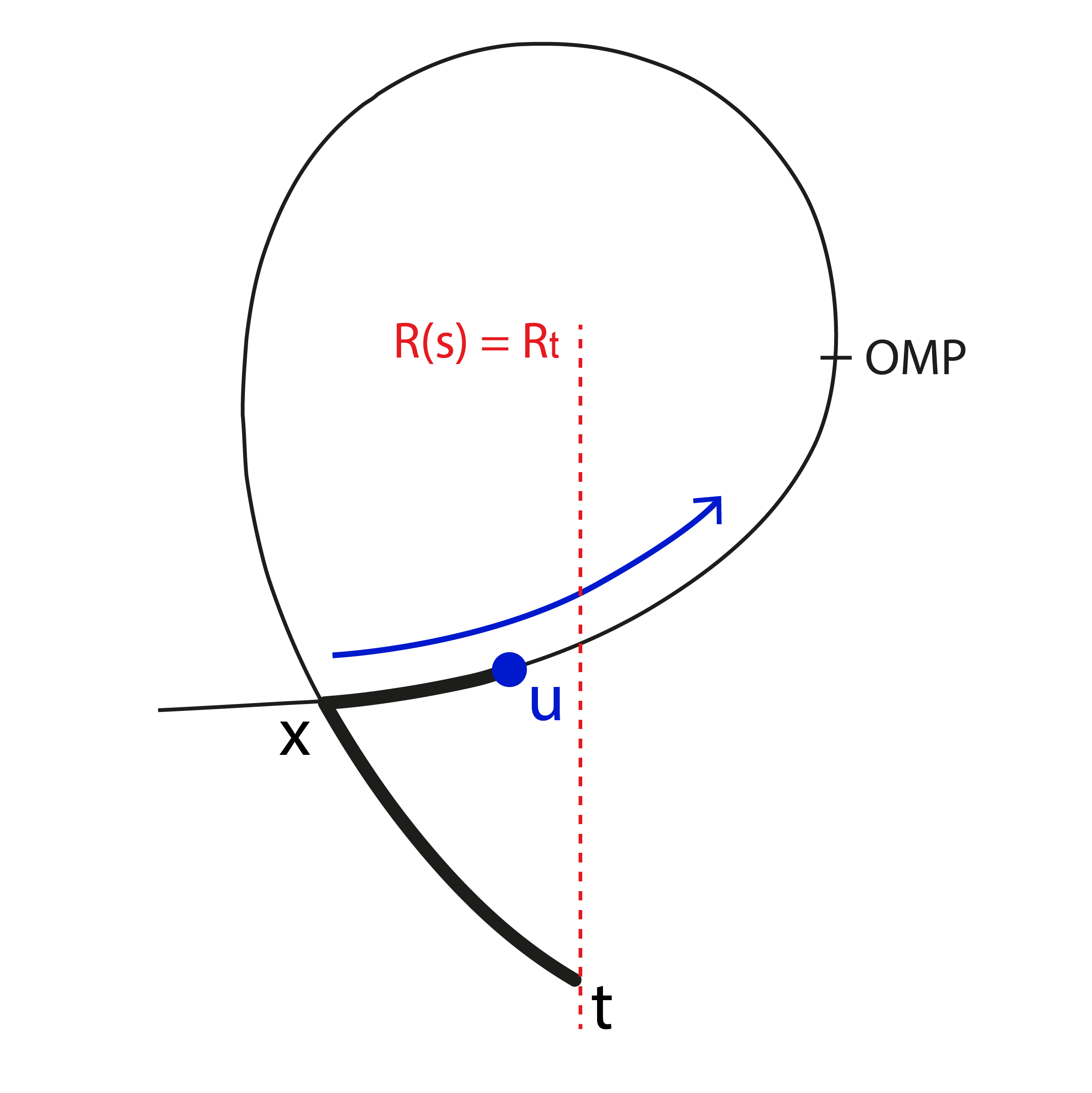}%
    
      \includegraphics[clip,width=0.86\columnwidth]{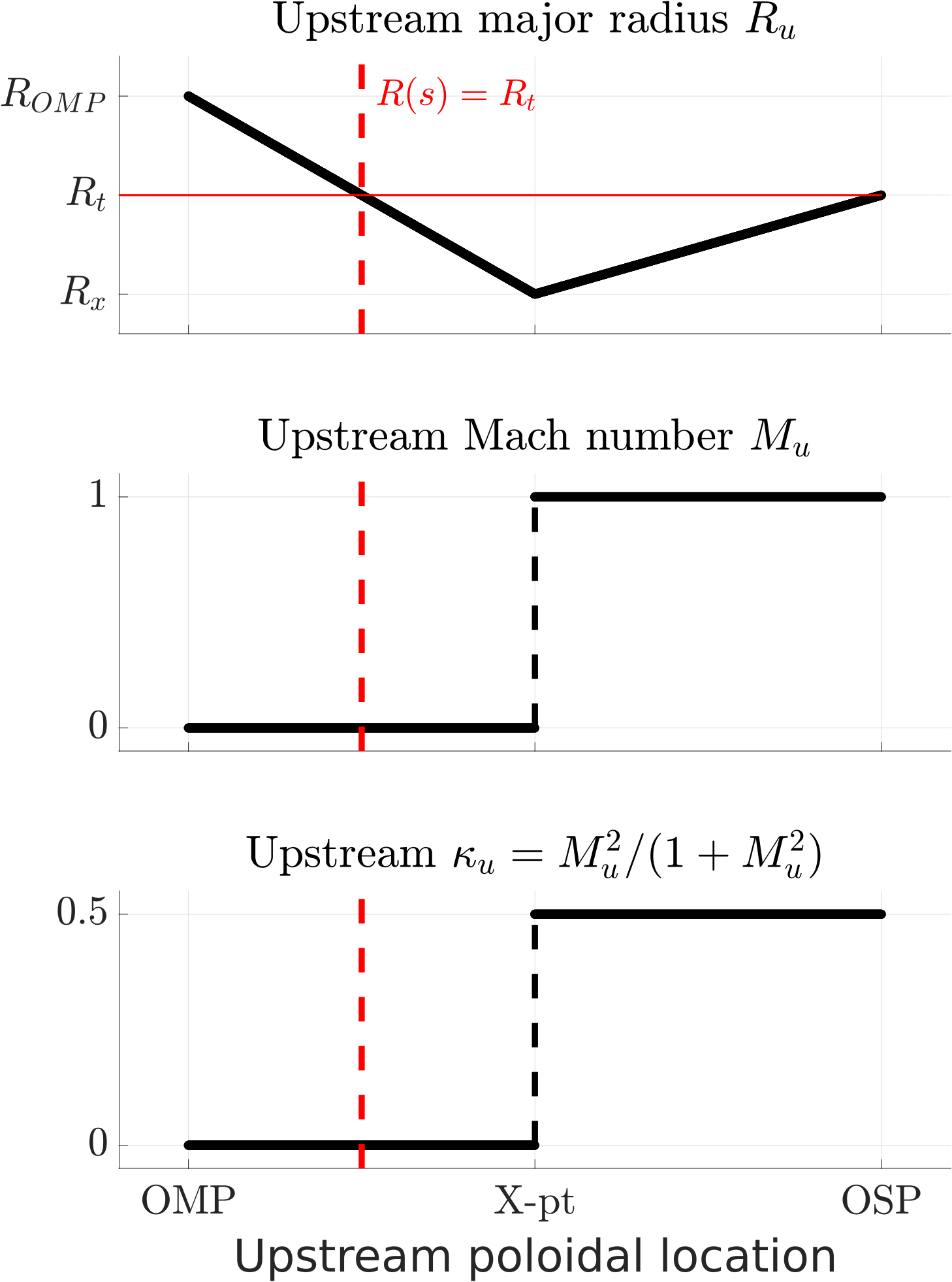}
    \caption{Graphical illustration of the example taken into consideration and variation of major radius $R$, Mach number $M$ and $\kappa = M^2/(1+M^2)$ against upstream poloidal location. The vertical dotted red line indicates the location at which $R=R_t$, between OMP and X-point.}
    \label{fig:pathological-exs2}
\end{figure}

\begin{figure}[h]
    \centering
    \includegraphics[width = 0.86\columnwidth]{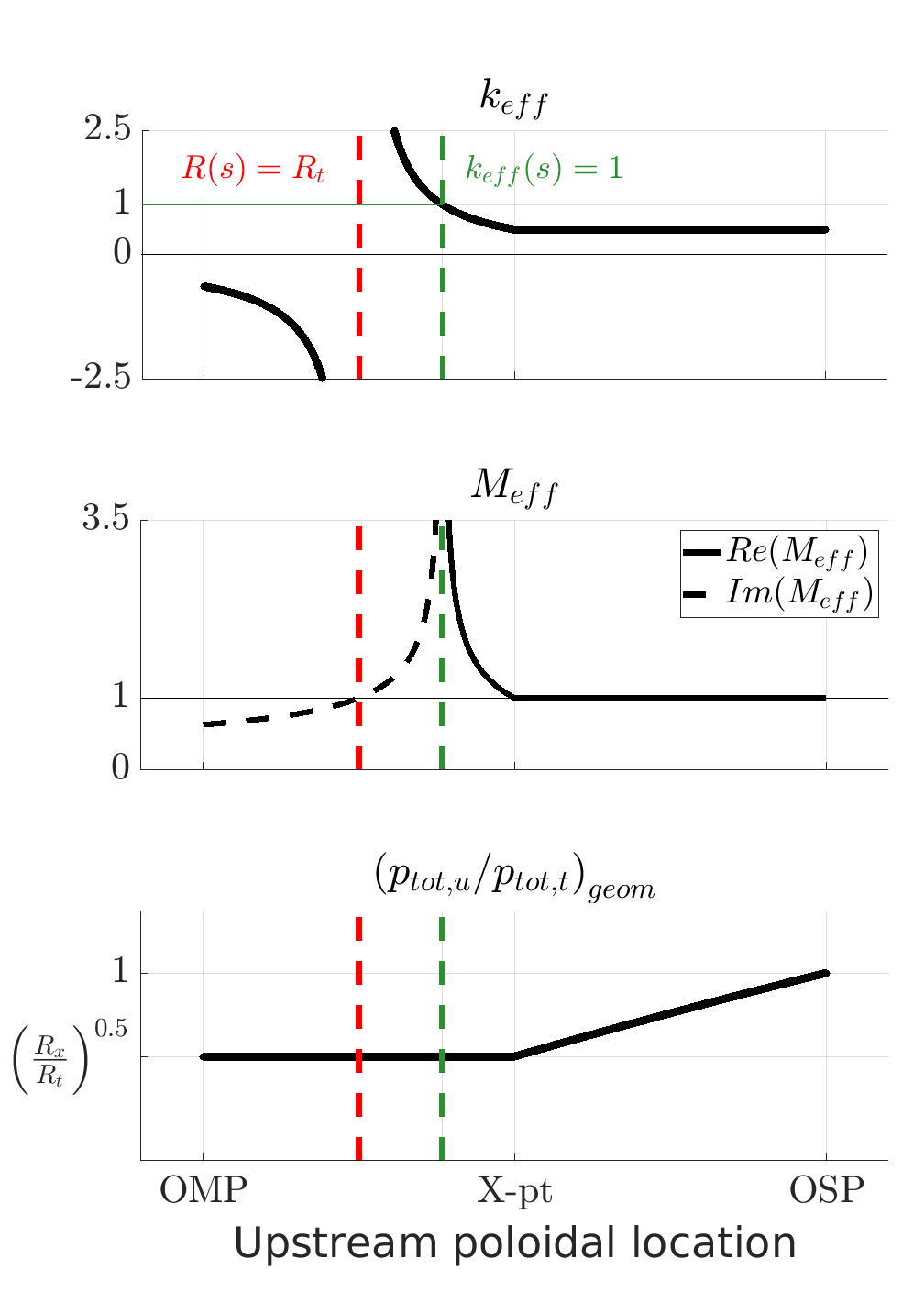}
    \caption{Variation of $\kappa_{eff}$, effective Mach number $M_{eff}$ and geometrical factor in the total pressure variation $(p_{tot,u}/p_{tot,t})_{geom}$ against upstream poloidal location in the example taken into consideration. The vertical dotted red line indicates the location at which $R=R_t$, between OMP and X-point. The vertical green line indicates the location at which $k_{eff}=1$.}
    \label{fig:pathological-res2}
\end{figure}

\begin{figure}[h]
    \centering
    \includegraphics[width = 0.91\columnwidth]{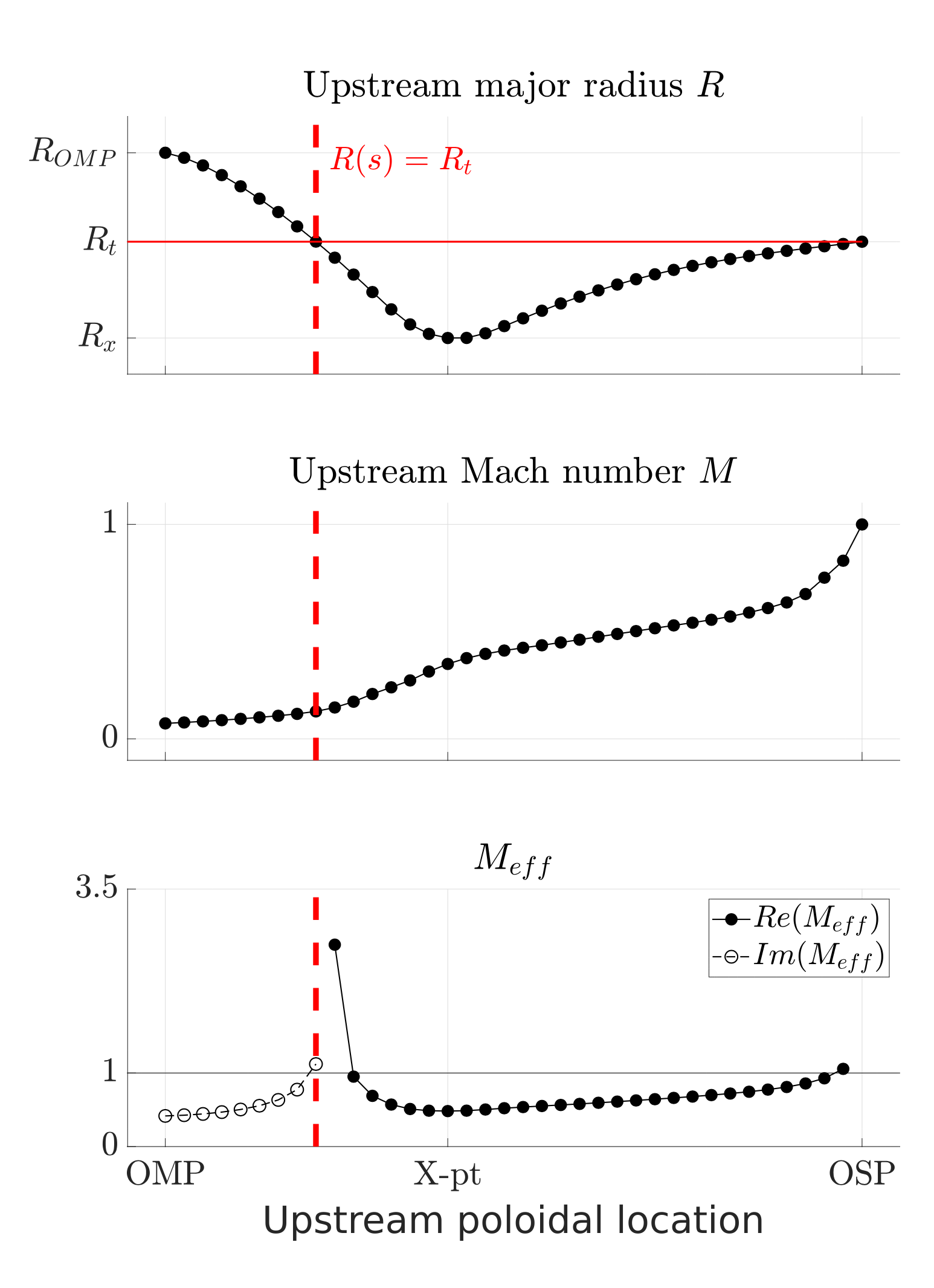}
    \caption{Variation of major radius $R$, Mach number $M$ and effective Mach number $M_{eff}$ against upstream poloidal location in the simulation presented in \ref{subsec:SOLPSmodel}. The intermediate fuelling case and the flux tube with the highest target temperature are considered. The vertical dotted red line indicates the location at which $R=R_t$, between OMP and X-point.}
    \label{fig:pathological-SIM2}
\end{figure}

Figure \ref{fig:pathological-exs2} shows a graphical visualisation of this example. Despite the $R$ variation, the total pressure $p_{tot}$ does not vary due to total flux expansion where $M=0$, \textit{i.e.} between the OMP and the X-point. $p_{tot}$ then gradually decreases, due to total flux expansion, between the X-point and the OSP, as $M=1$ and $R$ increases (see \eref{id-mombal2}). In this simple case, $\kappa_{eff}$ can be computed analytically for varying upstream location $s_u \in [s_{OMP}, s_t]$ by using \eref{keff-def}
\begin{eqnarray}
    k_{eff} = 0.5 \cdot \frac{\ln\left(R_t/R_x\right)}{\ln\left(R_t/R_u\right)} ~~& \mbox{ for } s_u \in [s_{OMP}, s_x) \nonumber \\
    k_{eff} = 0.5 & \mbox{ for } s_u \in [s_x, s_t] \nonumber
\end{eqnarray}

$M_{eff}$ can be then computed by \eref{Meff-def}, together with the geometrical factor in the total pressure variation $p_{tot,t}/p_{tot,u}$ (see \eref{fmom-R})
\begin{eqnarray}
    \left(\frac{p_{tot,t}}{p_{tot,u}}\right)_{geom} &= \left(\frac{R_u}{R_t}\right)^{0.5\cdot\ln(R_t/R_x)/\ln(R_t/R_u)} =\nonumber \\
    ~ &= e^{-0.5\cdot\ln(R_t/R_x)} = \nonumber \\ 
    ~ &= \left(\frac{R_x}{R_t}\right)^{0.5} ~~ \mbox{ for } s_u \in [s_{OMP}, s_x) \nonumber \\
    \left(\frac{p_{tot,t}}{p_{tot,u}}\right)_{geom} & = \left(\frac{R_u}{R_t}\right)^{0.5} ~~ \mbox{ for } s_u \in [s_x, s_t] \nonumber
\end{eqnarray}

The results are represented in figure \ref{fig:pathological-res2}.  For a choice of upstream location between the OSP and the X-point, where $M=1$, $M_{eff}$ and $k_{eff}$ are constants. The total pressure variation, due to total flux expansion, is reflected in the variation in $(p_{tot,u}/p_{tot,t})_{geom}$. When the upstream location is shifted beyond the X-point and towards the OMP, as the total pressure no longer varies due to total flux expansion, $(p_{tot,u}/p_{tot,t})_{geom}$ is constant. However, as $(R_u/R_t)$ keeps varying in this region (in this example, increasing towards the OMP), $M_{eff}$ also varies to accommodate this change. When $(R_u/R_t)$ increases above a given threshold (where $\kappa_{eff}=1$), a positive $M_{eff}$ can no longer accommodate this variation and imaginary values are obtained. This is understandable in terms of the definition provided above: taking for example the OMP as the upstream location, for which $R_u/R_t > 1$, there exists no constant value of $M \in \mathbb{R}$ which would result in a total pressure decrease towards the target, as in this example. 

Similar results can be obtained in more realistic cases, such as for example the SOLPS-ITER simulation analysed in section \ref{subsec:SOLPSmodel}. Also in this case, flux tubes feature a convergent-divergent magnetic structure, between the OMP and the OSP, and a monotonically increasing $M$ towards the OSP, figures \ref{fig:Meff-sim}\textcolor{blue}{a} and \ref{fig:pathological-SIM2}. These conditions tend to push the minimum for $M_{eff}$ close to the poloidal location where $R$ is minimum, that is often the X-point location for the standard geometry of outer legs in diverted configurations, figure \ref{fig:pathological-SIM2}. This justifies why the choice of the divertor entrance, as the upstream location to evaluate $M_{eff}$ in section \ref{subsec:SOLPSmodel}, was termed as conservative. 

\bigskip

\noindent \textit{Dependence of $M_{eff}$ on the divertor leg geometry}

\bigskip

$M_{eff}$ is derived, through $\kappa_{eff}$, from a weighted average of $\kappa = M^2/(1+M^2)$ along the flux tube, where the weighting factor is the local relative variation of the flux tube area $R^{-1}\partial_s(R)$ (see \eref{keff-def} - \eref{Meff-def}). This implies that for a given $M$ distribution, from upstream to target, the local flux expansion distribution along the leg influences the value of $M_{eff}$ and, therefore, the magnitude by which total flux expansion effects are reflected on total pressure variation, target quantities and detachment window. In other words, the divertor leg geometry influences the sensitivity to total flux expansion effects.

\begin{figure}[h!]
    \centering
    \includegraphics[width = 0.95\columnwidth]{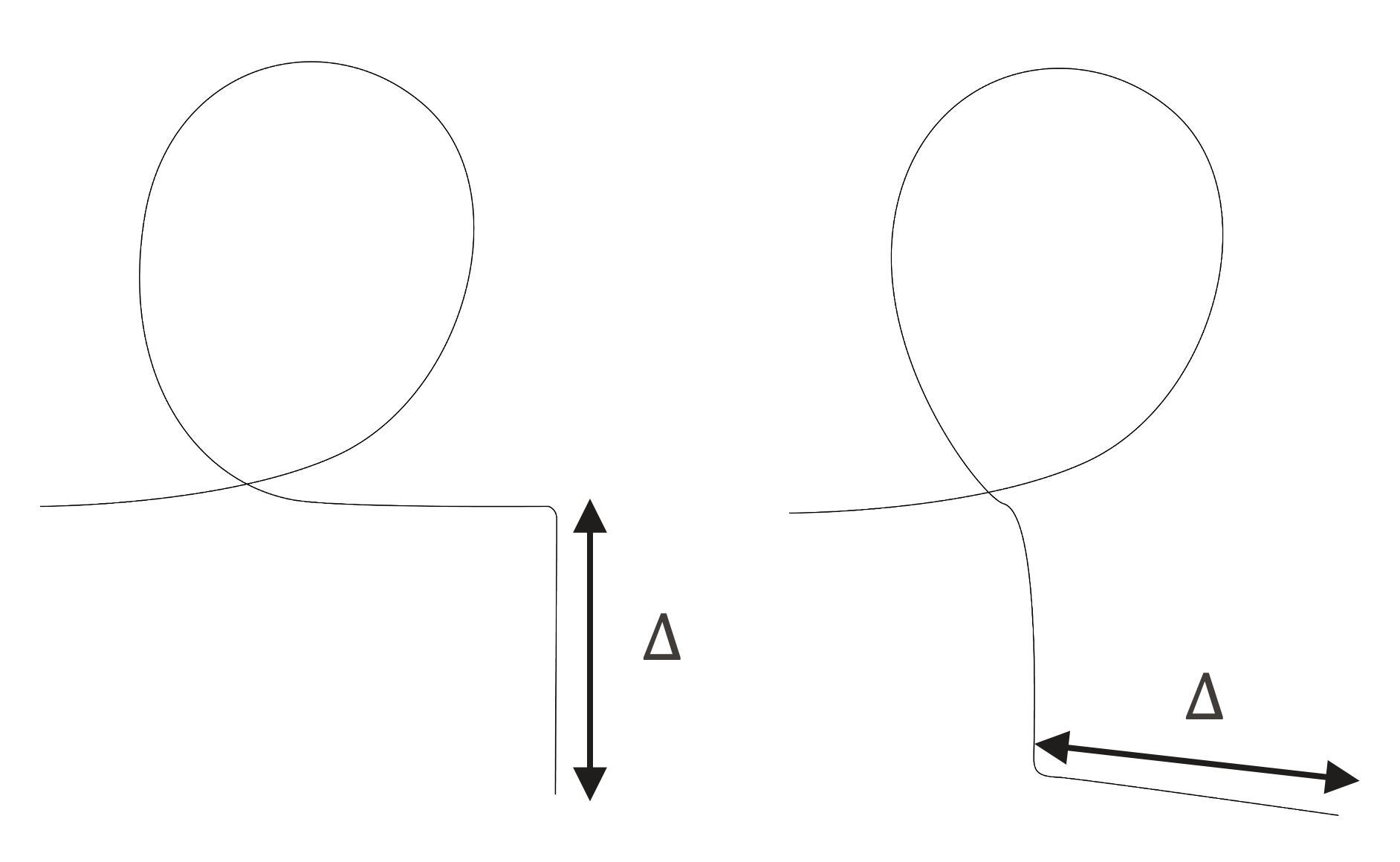}
    \caption{Graphical illustration of the two examples taken into consideration (case A on the left, case B on the right). Upstream location is here the X-point and the flux tube of interest is in the outer leg.}
    \label{fig:pathological-exs}
\end{figure}

Here, a couple of pathological examples are provided to better highlight this point. Two cases, with the same total flux expansion $R_t/R_u$, are considered in which the local flux expansion is constant and focused only: (case A) in the region where $M=0$; (case B) in the region where $M=1$. Consider a SOL flux tube and a field aligned length coordinate $s=[0,L]$, where $s=0$ corresponds to the upstream position (with major radius $R_u$) and $s=L$ corresponds to the target position (with major radius $R_t$). Assume the following profiles for the Mach number along the flux tube
\begin{eqnarray}
    M = 1 \cdot \chi[L-\Delta,L] \nonumber
\end{eqnarray}
and for local relative flux expansion
\begin{eqnarray}
    \mbox{(case A)}& ~~\frac{1}{R}\partial_s(R) = \ln{\frac{R_t}{R_u}} \cdot \frac{\chi[0,L-\Delta]}{(L-\Delta)} \nonumber \\
    \mbox{(case B)}& ~~\frac{1}{R}\partial_s(R) = \ln{\frac{R_t}{R_u}} \cdot \frac{\chi[L-\Delta,L]}{\Delta} \nonumber
\end{eqnarray}
where $\chi[s_1,s_2]$ is a function which equals 1 in between $s_1$ and $s_2$ and 0 elsewhere, and $\Delta\in(0,L)$. In practice, it is imposed that $M$ will increase instantaneously from $0$ to $1$ in the portion $[L-\Delta,L]$ of the flux tube in front of the target. Notice that in both cases the total flux expansion $R_t/R_u$ is the same. Figure \ref{fig:pathological-exs} shows a graphical visualisation of these examples, for the outer leg of a single-null configurations (taking the X-point as upstream).

Computing now $\kappa_{eff}$ by \eref{keff-def} and $M_{eff}$ by \eref{Meff-def}, it is found 
\begin{eqnarray}
    \mbox{(case A)}~~~ \kappa_{eff} = 0 ~ & \rightarrow ~~~ M_{eff} = 0 \nonumber \\
    \mbox{(case B)}~~~ \kappa_{eff} = 0.5 ~ & \rightarrow ~~~ M_{eff} = 1 \nonumber
\end{eqnarray}
It is then clear the drastic change in $M_{eff}$ depending on the geometry of the flux tube, considering the same total flux expansion and flows pattern.

\section[]{Derivation of detachment window expression}
\label{app:detwinder}

The derivation of \eref{detwin-1} is presented. This is similar to the original derivation reported in \cite{Lipschultz2016}. In addition, the same hypothesis of thermal equilibration in the flux tube (\textit{i.e.} $T=T_e=T_i$) is adopted, as in appendix \hyperref[app:2PMder]{A}. Therefore, the plasma static pressure is $p=2nT$. 

Consider the total steady-state energy balance in a flux tube. Assume (a) cross-field transport effects are negligible. Assume also that (b) the ratio $f_{cond}$ of conducted to total parallel power density is constant, and (c) Spitzer's formulation for parallel heat conductivity can be used: $\kappa_\parallel = \kappa_0 T^{5/2}$. The power balance is then
    
\begin{equation}
    H = - \frac{1}{f_{cond}} B \partial_s\left(\frac{\kappa_\parallel}{B}\partial_s(T)\right)
    \label{powerbalOG}
\end{equation}
    
where $s$ is the length coordinate along a field line from target to upstream, here considered as corresponding to the X-point ($s: ~[0,s_x]$).
    
It is assumed that (d) $H=-n^2f_IQ(T)$, which means the local effective power sources/sinks can be approximated with their only radiation-related component. Here $n$ is the plasma density, $f_I$ is the impurity fraction ($f_I=n_I/n$) and $Q(T)$ is a radiation efficiency function. The radiation efficiency $Q(T)$ is assumed (e) to be a function which peaks sharply just in a range $[T_c,T_h]$ (with $T_c < T_h$) and it's null outside of it.
    
The following change of variable is introduced

\begin{equation}
    dz = \frac{B_x}{B}ds
\end{equation}
    
Practically, $z=\int_0^z dz'=\int_0^{s(z)} \frac{B_x}{B}ds'$ will be the volume ($ds/B \propto dV$) of the flux tube contained from the target ($s,z=0$) up to the point of interest, normalized by a reference perpendicular area ($\propto 1/B_x$), where the upstream/X-point is taken as this reference.
    
Defining
    
\begin{equation}
    \kappa = \kappa_\parallel \frac{B_x^2}{B^2}
    \label{mod-heatres}
\end{equation}
    
\eref{powerbalOG} becomes
    
\begin{equation}
    \partial_z q = H
    \label{powerbalMOD}
\end{equation}
    
with 
    
\begin{equation}
    q = - \frac{1}{f_{cond}}\kappa \partial_z T
    \label{qdef}
\end{equation}
    
$q = (1/f_{cond})q_{\parallel,cond}B_x/B$ is then the total parallel power $Q_\parallel \propto (1/f_{cond})q_{\parallel,cond} / B$ normalized by the same reference perpendicular area $\propto 1/B_x$.
    
Taking \eref{powerbalMOD} and multiplying both sides by $q$, then integrating from $z(T_c)$ to $z(T_h)$ (note that $z(T_h)>z(T_c)$, in the chosen coordinate system)
    
\begin{equation}
    [q^2]_{z(T_c)}^{z(T_h)} = - \int_{T_c}^{T_h} \frac{2}{f_{cond}}\kappa(T')H(T')dT'
    \label{q2}
\end{equation}
    
Using assumptions (b)-(e), the square root of the integral on the right hand side of this equation becomes
    
\begin{equation}
    \Delta q_{rad} \equiv \sqrt{\frac{2\kappa_0}{f_{cond}} \int_{T_c}^{T_h} \frac{B_x^2}{B^2} T^{5/2} n^2 f_I Q(T) dT}
    \label{RadQ}
\end{equation}
    
Assume (f) the radiation region (\textit{i.e.} the region in between $z(T_c)$ and $z(T_h)$) is so narrow that $B$ and $f_I$ variations are negligible in it. Assuming also that (g) volumetric processes and cross-field transport effects on momentum balance and total pressure redistribution are negligible in this region, this implies $p^2=4n^2T^2=p_{tot}^2/(1+M^2)^2$ can be taken out of the integral as its variation will be then linked just to total flux expansion effects (hence, $B$ variation), negligible by assumption (f). Therefore
    
\begin{equation}
    \Delta q_{rad} = \frac{B_x}{B_{z(T_h)}} p_{z(T_h)} \sqrt{\frac{\kappa_0}{2f_{cond}} f_I \mathcal{F}}
    \label{RadQ2}
\end{equation}
  
with $\mathcal{F}=\int_{T_c}^{T_h}\sqrt{T} Q(T) dT$.
    
The pressure at the detachment front entrance $p_{z(T_h)}$ is linked with pressure upstream/at the X-point $p_u$ using \eref{fmom-R}, substituting $B\propto R^{-1}$. It is assumed that (h) volumetric processes and cross-field transport effects on momentum balance are negligible in the region between the X-point and the detachment front entrance. It then holds

\begin{eqnarray}
    \frac{p_{tot,z(T_h)}}{p_{tot,x}} & = \frac{1+M^2_{z(T_h)}}{1+M_x^2} \frac{p_{z(T_h)}}{p_{u}} = \\ & = \left( \frac{B_{z(T_h)}}{B_{x}} \right)^{\kappa_{eff}^{x \rightarrow z(T_h)}} \nonumber
\end{eqnarray}
    
\eref{RadQ2} becomes then
    
\begin{eqnarray}
    \Delta q_{rad} & = & \frac{B_{x}}{B_{z(T_h)}} \frac{1 + M_x^2}{1+M^2_{z(T_h)}} \\ & ~ & \cdot \left(\frac{B_{z(T_h)}}{B_{x}}\right)^{\kappa_{eff}^{x \rightarrow z(T_h)}} p_u \sqrt{\frac{\kappa_0}{2f_{cond}} f_I \mathcal{F}} = \nonumber \\
    ~ & = & \frac{1 + M_x^2}{1+M^2_{z(T_h)}} \left( \frac{B_{x}}{B_{z(T_h)}} \right)^{1-\kappa_{eff}^{x \rightarrow z(T_h)}} \nonumber \\ & ~ & \cdot p_u \sqrt{\frac{\kappa_0}{2f_{cond}} f_I \mathcal{F}} \nonumber
    \label{RadQ4}
\end{eqnarray}
    
Finally, to obtain a model for the operational window for different control parameters, it is assumed that (i) the power leaving the cold detachment front is negligible. This will imply $q_{z(T_h)} = - \Delta q_{rad}$ by \eref{q2}. The power entering the hot detachment front must then match the power entering upstream/at the X-point, thanks to assumption (d), and the latter can be expressed as $q_i = - P_{SOL}$, by definition of $q$. 

Now one can equate $q_{z(T_h)}$ and $q_i$ and solve in terms of the control parameters $\zeta = [p_u, f_I,P_{SOL}]$. The front position $z(T_h)$ is then set to be at the X-point first and then at the target to find the corresponding values $\zeta_{x,t}$ (leaving the others parameters constant). Dividing these two values, the detachment window is obtained
    
\begin{equation}
    \frac{\zeta_x}{\zeta_t} = \left( \left(\frac{B_{tot,x}}{B_{tot,t}}\right)^{1-k_{eff}} \frac{1+M_x^2}{1+M_t^2}~\right)^\beta
\end{equation}

with $\beta = [1,2,-1]$.

\section[]{Derivation of Mach number evolution equation}
\label{app:MachEvo}

Consider the steady-state ion particle balance and plasma momentum balance along a flux tube

\begin{eqnarray}
    B \partial_s \left(\frac{nv_s}{B}\right) & = S_{par} \label{eq-parbalIV} \\
    B \partial_s \left(\frac{m_inv^2_s}{B}\right) & = - \partial_s(nT^*) + S_{mom} \label{eq-mombalIV}
\end{eqnarray}

where $s$ is a length reference coordinate along the flux tube and $S_{par,mom}$ includes contributions from volumetric sources and cross-field transport effects. A single hydrogenic ion species and quasi-neutrality ($n_e=n_i=n$) are considered. For the sake of simplicity in the notation, $T^*=T_e+T_i$ is introduced.

Start rewriting the pressure term in \eref{eq-mombalIV}
\begin{eqnarray}
        B \partial_s \left(\frac{m_inv^2_s}{B}\right) = & - B\partial_s\left(\frac{nT^*}{B}\right ) \label{eq-mombalIV2} \\ & - \frac{nT^*}{B}\partial_s(B) + S_{mom} \nonumber
\end{eqnarray}
In both \eref{eq-parbalIV} and \eref{eq-mombalIV2}, isolate $\partial_s(n)$
\begin{eqnarray}
    \partial_s(n) = \frac{S_{par}}{v_s} - \frac{nB}{v_s}\partial_s \left( \frac{v_s}{B} \right ) \label{eq-parbalIV3} \\
    m_iv_s^2\partial_s(n) + nB \partial_s\left(\frac{mv_s^2}{B}\right) = - T^*\partial_s(n) \label{eq-mombalIV3} \\ ~~~~~~~~~~~ - nB\partial_s\left(\frac{T}{B}\right) - \frac{nT^*}{B}\partial_s(B) + S_{mom} \nonumber
\end{eqnarray}
Reordering and inserting \eref{eq-parbalIV3} into \eref{eq-mombalIV3}
\begin{eqnarray}
    -\frac{nB}{v_s}(m_iv_s^2 + T^*)\partial_s \left( \frac{v_s}{B} \right) + (m_iv_s^2 + T^*)\frac{S_{par}}{v_s} = \nonumber \\
    = -nB\partial_s\left( \frac{m_iv_s^2 + T^*}{B}\right) \nonumber \\ ~~~ - \frac{nT^*}{B}\partial_s(B) + S_{mom} \label{eq-mombalIV4}
\end{eqnarray}
Introducing $c_s = \sqrt{T^*/m_i}$ and reordering
\begin{eqnarray}
    -B(v_s^2+c_s^2)\partial_s\left(\frac{v_s}{B}\right) + v_sB\partial_s\left(\frac{v_s^2+c_s^2}{B}\right) = \nonumber \\
    = -(v_s^2+c_s^2)\frac{S_{par}}{n}-v_s\frac{c_s^2}{B}\partial_s(B) + \frac{v_sS_{mom}}{m_in}
    \label{eq-mombalIV5}
\end{eqnarray}
The left-hand-side of this equation is equivalent to 
\begin{equation}
    -(c_s^2-v_s^2)\partial_s(v_s) + 2v_sc_s\partial_s(c_s)
\end{equation}
Exploiting this and introducing $M = v_s/c_s$, one obtains
\begin{eqnarray}
    \frac{1-M^2}{c_s}\partial_s(v_s) = 2\frac{M}{c_s}\partial_s(c_s)  \label{eq-mombalIV6} \\~~~~ + (1+M^2)\frac{S_{par}}{nc_s} + \frac{M}{B}\partial_s(B)-\frac{MS_{mom}}{m_inc_s^2}
    \nonumber
\end{eqnarray}
which can be rewritten as
\begin{equation}
    \partial_s(v_s) = \partial_s(Mc_s) = M\partial_s(c_s) + c_s\partial(M)
\end{equation}
Exploiting this in \eref{eq-mombalIV6} and using $B \propto (A_\perp)^{-1}$, it is finally possible to retrieve \eref{eq-MachEvo}
\begin{eqnarray}
    (1-M^2) \partial_s (M) = & \frac{1+M^2}{nc_s}S_{par} \\ ~ & + \frac{M(1+M^2)}{c_s}\partial_s(c_s) \nonumber \\
    ~ & + A_\perp M \partial_s(\frac{1}{A_\perp}) \nonumber \\ ~ & - \frac{M}{m_inc_s^2}S_{mom} \nonumber
\end{eqnarray}

\section*{References}
\bibliographystyle{iopart-num}
\bibliography{main}

\end{document}